\documentstyle{article}
\begin{document}

\title{Expressing Products of Fermi Fields in terms of Fermi Sea Displacements}
\author{Girish S. Setlur}
    
\maketitle

\begin{abstract}                   
 An attempt is made to generalise the ideas introduced by Haldane and others
 regarding Bosonizing the Fermi surface. The present attempt involves
 introduction of Bose fields that correspond to displacements of the Fermi
 sea rather than just the Fermi surface. This enables the study of
 short wavelength fluctuations of the Fermi surface and hence
 the dispersion of single particle excitations with high energy.
 The number conserving product
 of two Fermi fields is represented as a simple combination of these
 Bose fields. It is shown that most(!) commutation rules involving these
 number conserving products are reproduced exactly,
 as are the dynamical correlation functions of the free theory.
 Also the work of Sharp, Menikoff and Goldin has shown that the field
 operator may be viewed as a Unitary  representation of the 
 local current group. An explicit realisation of this
 unitary representation is
 given in terms of canonical conjugate of the density operator.
\end{abstract}

\section{Introduction}

 Recent years have seen remarkable developments in many-body theory
 in the form an assortment of techniques that may be loosely termed
 bosonization. The beginnings of these types of techiniques may be
 traced back to the work of Tomonaga\cite{Tom}
 and later on by Luttinger \cite{Lutt} and by Leib and Mattis 
 \cite{Leib}. 
 In the 70's an attempt was made by Luther \cite{Luther}
 at generalising these ideas to higher dimensions. Closely related to this
 is work by Sharp et. al. \cite{Sharp} in current algebra. Mention must
 also be made about the work of Feenberg and his collaborators\cite{Feen}
 who have a theory of correlated electrons written down in the standard
 wavefunction approach familiar in elementary quantum mechanics. It seems
 that the approach adopted here has some similarites with this work
 although the details are not identical.
 More progress was made by Haldane \cite{Haldane} which culminated in
 the explicit computation of the single particle propagator by 
 Castro-Neto and Fradkin \cite{Neto} and by Houghton, Marston et.al.
 \cite{Mars} and also by Kopietz et. al. \cite{Kop}.
 Rigorous work by Frohlich and Marchetti\cite{Froch}
 is also along similar lines. 

 The attempt made here is to generalise the concepts of Haldane \cite{Haldane}
 to accomodate short wavelengths fluctuations where the concept of
 a linerised bare fermion energy dispersion is no longer valid.
 To motivate progress in this direction one must first introduce the concept of
 the canonical conjugate of the Fermi density distribution. This concept is
 likely to be important in the construction of the field operator in terms of
 the Fermi sea displacements, although the latter is not completed in this
 article. The concept of the velocity operator being the canonical
 conjugate of the density has been around for a long time, and this
 has been exploited in the study of HeII by Sunakawa et. al. \cite{Sun}.
 However, the author is not aware of a rigorous study of the meaning
 of this object, in particular, an explicit formula for the
 canonical conjugate of the density operator has to the best of my
 knowledge never been written down in terms of the field operators.
 The work by Sharp et. al. \cite{Sharp} comes close to what I am
 attempting here. 

\section{Unitary Representation of the Local Current Group}

 The work of Sharp et. al. \cite{Sharp} established that the current algebra
 itself does not convey the underlying particle statistics
 rather the statistics is hidden in a choice of a unitary
 representation of the local current group. Here, I try to write down
 some formulas that provide (possibly for the first time) an explicit
 and exact representation of the Fermi field operator in terms
 of the canonical conjugate of the density operator thereby providing
 an explicit realisation of the claims of Sharp et. al. \cite{Sharp}
\subsection{Some Mathematical Identities}

\begin{equation}
 \rho({\bf{x}}\sigma) = \psi^{\dagger}({\bf{x}}\sigma)\psi({\bf{x}}\sigma)
\end{equation}
is a bose-like object; namely it satisfies
\begin{equation}
 [\rho({\bf{x}}\sigma) , \rho({\bf{x^{'}}}\sigma^{'})] = 0
\end{equation}
Notation:
\[
N_{\sigma} = \int d{\bf{x}} \mbox{ }\rho({\bf{x}}\sigma)
\]
\[
n_{\sigma} = \frac{N_{\sigma}}{V}
\]
\[
\rho_{\sigma} = \langle n_{\sigma} \rangle
\]
$ \rho_{\sigma} $ is a c-number, whereas
$ N_{\sigma} $ and $ n_{\sigma} $ are operators.
                                                            
So it natural to introduce the canonical conjugate $ \Pi({\bf{x}}\sigma) $.
\begin{equation}
 [\Pi({\bf{x}}\sigma), \rho({\bf{x^{'}}}\sigma^{'})] =
i\delta^{d}({\bf{x-x^{'}}})\delta_{\sigma,\sigma^{'}}
\end{equation}
\begin{equation}
 [\Pi({\bf{x}}\sigma), \Pi({\bf{x^{'}}}\sigma^{'})] = 0
\end{equation}
Propose the ansatz
\begin{equation}
\psi({\bf{x}}\sigma) = exp(-i\Pi({\bf{x}}\sigma))
exp(i\Phi([\rho];{\bf{x}}\sigma)) (\rho({\bf{x}}\sigma))^{\frac{1}{2}}
\label{DPVA}
\end{equation}
The above Eq.(~\ref{DPVA}) shall be called the DPVA ansatz.
Here again,
\begin{equation}
\Pi({\bf{x}}\sigma) =
 i\ln[ (\rho_{\sigma}^{\frac{1}{2}}+\delta\psi({\bf x}\sigma))
(\rho_{\sigma} + \delta\rho({\bf{x}}\sigma))^{-\frac{1}{2}}
exp(-i\Phi([\rho];{\bf{x}}\sigma)) ]
\label{Rep}
\end{equation}
\begin{equation}
\delta\psi({\bf x}\sigma) = \psi({\bf x}\sigma) - \rho_{\sigma}^{\frac{1}{2}}
\end{equation}
\begin{equation}
\delta\rho({\bf{x}}\sigma) = \rho({\bf{x}}\sigma) - \rho_{\sigma}
\end{equation}
$ \rho_{\sigma} $ is a c-number given by
 $ \rho_{\sigma} = \langle  n_{{\sigma}} \rangle $.
Where $ \Phi([\rho];{\bf{x}}\sigma) $ is some hermitian functional
to be computed later.
This ansatz automatically satsfies
\begin{equation}
\psi^{\dagger}({\bf{x}}\sigma)\psi({\bf{x}}\sigma) = \rho({\bf{x}}\sigma)
\end{equation}
and
\begin{equation}
[\psi({\bf{x}}\sigma), \rho({\bf{x^{'}}}\sigma^{'})] =
 \delta^{d}({\bf{x - x^{'}}}) \delta_{\sigma, \sigma^{'}} \psi({\bf{x}}\sigma)
\end{equation}
Write
\begin{equation}
\rho({\bf{x}}\sigma) = n_{\sigma}
+ \frac{1}{V}\sum_{{\bf{q}}\neq 0} \rho_{{\bf{q}}\sigma} exp(-i{{\bf{q.x}}})
\end{equation}
where $  n_{\sigma} = \frac{N_{\sigma}}{V}  $.
and
\begin{equation}
\Pi({\bf{x}}\sigma) = X_{{\bf{0}}\sigma} +
 \sum_{{\bf{q}}\neq 0} exp(i{{\bf{q.x}}})
X_{{\bf{q}}\sigma}
\end{equation}
\begin{equation}
[X_{{\bf{0}}\sigma}, N_{\sigma^{'}}] = i \delta_{\sigma, \sigma^{'}}
\end{equation}
\begin{equation}
[X_{{\bf{q}}\sigma}, \rho_{{\bf{q^{'}}}\sigma^{'}}] =
i \delta_{{\bf{q}},{\bf{q^{'}}}} \delta_{\sigma, \sigma^{'}}
\end{equation}
Therefore
\begin{equation}
\psi({\bf{x}}\sigma) = exp(-i \sum_{{\bf{q}}} exp(i{{\bf{q.x}}})
X_{{\bf{q}}\sigma} ) F([\{ \rho_{{\bf{k}}\sigma^{'}} \}]; {\bf{x}}\sigma)
\end{equation}
where
\begin{equation}
 F([\{ \rho_{{\bf{k}}\sigma^{'}} \}]; {\bf{x}}\sigma) =
exp(i\Phi([\{ \rho_{{\bf{k}}\sigma^{'}} \}]; {\bf{x}}\sigma)) (n_{\sigma}
+ \frac{1}{V}\sum_{{\bf{q}}\neq 0} exp(-i{{\bf{q.x}}}) \rho_{{\bf{q}}\sigma})
^{\frac{1}{2}}
\end{equation}
It is possible to write these formulas in momentum space. This will be useful
later on.
First we write
\begin{equation}
\Phi([\rho]; {\bf{x}}\sigma) = \sum_{{\bf{q}}}
\phi( [\rho]; {\bf{q}}\sigma ) exp(-i{\bf{q.x}})
\end{equation}
Therefore
\begin{equation}
\psi({\bf{x}} \sigma) =
exp(-i \sum_{{\bf{q}}} exp(i{\bf{q.x}}) X_{{\bf{q}}\sigma} )
exp(i \sum_{{\bf{q}}} exp(-i{\bf{q.x}}) \phi( [\rho]; {\bf{q}}\sigma ) )
(n_{\sigma} + \frac{1}{V} \sum_{{\bf{q}}\neq 0}
 \rho_{{\bf{q}}\sigma} exp(-i{\bf{q.x}}) )^{\frac{1}{2}}
\end{equation}
It is possible to write the corresponding formula in momentum space
by making the identifications
\begin{equation}
exp(i{\bf{q.x}}) \rightarrow T_{{\bf{-q}}}({\bf{k}})
\end{equation}
\begin{equation}
exp(-i{\bf{q.x}}) \rightarrow T_{{\bf{q}}}({\bf{k}})
\end{equation}
where
\begin{equation}
T_{{\bf{q}}}({\bf{k}}) = exp({\bf{q}}.\nabla_{{\bf{k}}})
\end{equation}
\begin{equation}
\psi({\bf{k}}\sigma) =
exp(-i \sum_{{\bf{q}}} T_{{\bf{-q}}}({\bf{k}})  X_{{\bf{q}}\sigma} )
exp(i \sum_{{\bf{q}}} T_{{\bf{q}}}({\bf{k}})\phi( [\rho]; {\bf{q}}\sigma ) )
(N_{\sigma} + \sum_{{\bf{q}}\neq 0}
 \rho_{{\bf{q}}\sigma} T_{{\bf{q}}}({\bf{k}}) )^{\frac{1}{2}}
\delta_{{\bf{k}}, 0}
\label{Fermi}
\end{equation}
The translation operators translate the Kronecker delta that appears in the
 extreme right.
It may be verified that
\begin{equation}
[\psi({\bf{k}}\sigma), \rho_{{\bf{q}}\sigma^{'}}] =
\psi({\bf{k-q}}\sigma)
\delta_{\sigma,\sigma^{'}}
\end{equation}
\begin{equation}
[\psi({\bf{k}}\sigma), N_{\sigma^{'}}] = \psi({\bf{k}}\sigma)
\delta_{\sigma,\sigma^{'}}
\end{equation}
Here
\begin{equation}
\psi({\bf{x}}\sigma) = \frac{1}{{V^{\frac{1}{2}}}}\sum_{{\bf{k}}}
exp(i{\bf{k.x}})\psi({\bf{k}}\sigma)
\end{equation}
Also the following identities are going to be important.
\begin{equation}
\psi({\bf{k}}\sigma) =
 exp( -i\sum_{\bf{q}}T_{ -{\bf{q}} }({\bf{k}})X_{ {\bf{q}} \sigma } )
 f_{{\bf{k}}\sigma}([\rho])
\end{equation}
\begin{equation}
f_{ {\bf{k}} \sigma}([\rho]) =
 exp( i\sum_{ {\bf{q}} }T_{ {\bf{q}} }({\bf{k}})
 \phi([\rho]; {\bf{q}} \sigma) )
(N_{ \sigma } + \sum_{ {\bf{q}}\neq 0 }
 T_{ {\bf{q}} }({\bf{k}}) \rho_{ {\bf{q}}\sigma })^{\frac{1}{2}}
\delta_{ {\bf{k, 0}} }
\end{equation}
Check by expansion that
\begin{equation}
\sum_{ {\bf{k}} }
f^{\dagger}_{ {\bf{k+q/2}} \sigma}([\rho])
 f_{ {\bf{k-q/2}} \sigma}([\rho])
 = \rho_{ {\bf{q}}\sigma }
\end{equation}
\subsection{Proof (??) of the Fermion Commutation Rules}
 We use the theory of distributions to ( try ) prove rigorously
  (good enough for a
 physicist!) the fermion anticommutation rules.
 Pardon the pretense of mathematical rigor.
 The approach is as follows.
 Let $   $ be the space of all smooth functions from
 $ S = R^{3} \times \{ \uparrow, \downarrow \} $ to $ C $. Further, on this
 space define the inner product (Schwartz space).
 $ In: L^{2}_{C}(S) \times  L^{2}_{C}(S) \rightarrow C $
\[
\langle f|g \rangle = \int d{\bf{x}} \sum_{\sigma}
 f^{*}({\bf{x}}\sigma) g({\bf{x}}\sigma)
\]
Since the Fermi fields are operator-valued distributions we can construct
 operators
\[
c_{f} = \int d{\bf{x}}\sum_{\sigma}\psi({\bf{x}}\sigma)f({\bf{x}}\sigma)
\]
and we then have to prove
\begin{equation}
\{ c_{f}, c_{g} \} = 0
\end{equation}
\begin{equation}
c_{f}^{2} = 0
\end{equation}
The above two relations shall be called $ c-c $ anticommutation rules.
\begin{equation}
\{ c_{f}, c^{\dagger}_{g} \} = \langle g|f \rangle
\end{equation}
The above relation shall be called the $ c-c^{\dagger} $ anticommutation rule.
for all $ f, g $ that belong to $ L^{2}_{C}(S) $.
The claim is, that these relations are satisfied provided $ \Phi $ obeys
the recursion below.
\[
\Phi([\{\rho({\bf{y_{1}}}\sigma_{1})
 - \delta({\bf{y_{1}}}-{\bf{x}}^{'})\delta_{\sigma_{1},\sigma^{'}} \} ]
;{\bf{x}}\sigma)
\]
\[
+ \Phi([\rho];{\bf{x^{'}}}\sigma^{'}) - \Phi([\rho];{\bf{x}}\sigma)
\]
\begin{equation}
-\Phi([\{\rho({\bf{y_{1}}}\sigma_{1})
 - \delta({\bf{y_{1}}}-{\bf{x}})\delta_{\sigma_{1},\sigma} \} ]
;{\bf{x^{'}}}\sigma^{'})
 = m\pi
\label{recur}
\end{equation}
where m is an odd integer. This recursion is to be satisfied for all
$ ({\bf{x}}\sigma) \neq  ({\bf{x^{'}}}\sigma^{'}) $. At precisely
 $ ({\bf{x}}\sigma) =  ({\bf{x^{'}}}\sigma^{'}) $ the recursion
 obviously breaks down. But this is not a serious drawback as we shall
 soon find out. The proof involves working with a basis.
 Here the integer $ m $ has to be even for bosons and odd for fermions.
 The imposition of the commutation rules by themselves do not
 provide us with a formula for the hermitian functional $ \Phi $.
 Any redefinition of $ \Pi({\bf{x}}\sigma) $ consistent with
 $ [\Pi, \rho] = i\delta(...) $ may be absorbed by a suitable
 redefinition of $ \Phi $ (akin to gauge transformations pointed out by
 AHC Neto, private communication).
 Solution to $ \Phi([\rho];{\bf{q}}\sigma) $ by making contact with the
 free theory. A random choice of $ \Phi([\rho];{\bf{x}}\sigma) $ that
 satisfies the recursion is (A. J. Leggett : private communication)
 the Leggett ansatz (for fermions) (Also F.D.M. Haldane \cite{Haldane}).
\[
\Phi([\rho];{\bf{x}}\sigma)  =
 \int \mbox{ }d{\bf{x^{'}}} \sum_{{\sigma^{'}}}
 \pi\theta(t({\bf{x}}\sigma) - t({\bf{x^{'}}}\sigma^{'}))
\rho({\bf{x^{'}}}\sigma^{'})
\]
 redefinitions of $ \Phi $ that leave the statistics invariant
 are the 'generalised guage transformations'.
 Disclaimer:Please pardon my pretensions at rigor in the following paragraph.
 It is easy to get carried away.
 The relation that
 relates two fermi fields by a generalised guage transformation is
 an equivalence relation. Therefore , analogous to the claim that
 a state of a system is a ray in Hilbert space, one is tempted to
 call the equivalence class of fermi fields under these transformations
 the fermi distribution. Therefore a fermi distribution is not just one
 fermi field but a whole bunch of equivalent ones. Therefore a state containing
 one fermion at a space point is obtained by acting the creation
 fermi distribution on the vacuum. Thus one may construct the Fock space
 by repeatedly acting these fermi distributions on the vacuum.
 Therefore there are two ways by which one can choose $ \Phi $.
 One is a random choice which satisfies the recursion. In which case
 $ \Pi $ is determined by Eq.(\ref{Rep}). The other choice is
 $ \Pi = i\frac{\delta}{\delta \rho} $. In which case, $ \Phi $ can no
 longer be chosen arbitrarily. It has to be determined by making contact
 with the free theory.

 Here $ \theta(x) $ is the Heaviside step function. The unpleasentness
 caused by the fact that the recursion for fermions does not hold
 when $ ({\bf{x}}\sigma) = ({\bf{x^{'}}}\sigma^{'}) $ is probably
 remediable by multiplying the fermi fields which are operator-valued
 distributions by arbitrary smooth functions and proving properties about these
 latter objects. More importantly, $ t $ has to invertible in order for
 the ansatz to satisfy the recursion. That such a bijective mapping exists is
 guaranteed by the theory of cardinals. However, this mapping is not
 continuous let alone differentiable and therefore of little practical value.
 (A continuous mapping would imply a homeomorphism between
 $ R $ and $ R^{3} $ eg.)
                                            
 The claim is that the state of the
 fermi system is prescribed by prescribing an amplitude for
 finding the system in a given configuration of densities.
 Let
\[
 W_{FS}([\rho]) = \Theta_{H}([\rho])exp(-U_{FS}[\rho])
exp(i\theta_{FS}([\rho]))
\]
 be the
 wavefunctional of the noninteracting fermi sea. The $ U_{FS}[\rho] $ is
 uniquely determined by prescribing all moments of the density operator.
 The functional $ \theta_{FS}[\rho] $ cannot be so determined.  Moreover, since
 the amplitude for finding the system with negative densities is zero,
 we must also have a prefactor,
\[
\Theta_{H}([\rho]) = \Pi_{{\bf{y_{1}\sigma_{1}}}}
\theta_{H}(\rho({\bf{y_{1}\sigma_{1}}}) )
\]

Where $ \theta_{H}(x) $ is the Heaviside unit step function.
 \subsection{Making Contact With the Free Theory}
 Consider the operator
\begin{equation}
n({\bf{y}}\sigma) =
\int d{\bf{x}}\mbox{ }
\psi^{\dagger}({\bf{x+y/2}}\sigma)\psi({\bf{x-y/2}}\sigma)
 = \sum_{{\bf{k}}}\psi^{\dagger}({\bf{k}}\sigma)\psi({\bf{k}}\sigma)
exp(-i{\bf{k.y}})
\end{equation}
\begin{equation}
n({\bf{y}}\sigma)W_{FS}([\rho]) = n_{0}({\bf{y}}\sigma)W_{FS}([\rho])
\label{wavef}
\end{equation}
\begin{equation}
n_{0}({\bf{y}}\sigma) = \sum_{{\bf{k}}}\theta(k_{F}-k)
exp(-i{\bf{k.y}})
\end{equation}
 First assume $ {\bf{y}} \neq {\bf{0}} $.  In eq. (~\ref{wavef}) there are two
 undetermined functionals. One is $ \Phi([\rho];{\bf{x}}\sigma) $
 and the other is $ \theta_{FS}([\rho]) $. We have remarked earlier that
 redefinitions of $ \Phi([\rho];{\bf{x}}\sigma) $ are similar to
'gauge transformations'. These functional gauge transformations leave
 the local density of particles invariant but alter the statistics.
 There is a subgroup of these functional gauge transformations that
 also leaves the statistics invariant. A realisation of this subgroup
 is achieved by for example:
\[
 \Phi([\rho];{\bf{x}}\sigma) \rightarrow
 \Phi([\rho];{\bf{x}}\sigma)+\theta([\rho])
\]
This may be exploited to our advantage while solving eq.(~\ref{wavef}). In other
 words $ \theta_{FS}([\rho]) $ can be absorbed into
 $ \Phi([\rho];{\bf{x}}\sigma) $ without altering the statistics.
This leaves us with an equation for $ \Phi([\rho];{\bf{x}}\sigma) $ in terms
 of the properties of the free theory exclusively. Exploiting the recursion
relation for $ \Phi([\rho];{\bf{x}}\sigma) $ we arrive at the result
(Here $ y \neq 0 $.)
\[
\int d{\bf{x}} \mbox{ } (\rho({\bf{(x+y/2)}}\sigma))^{\frac{1}{2}}
 (\rho({\bf{(x-y/2)}}\sigma))^{\frac{1}{2}}
exp[i\Phi([\rho];{\bf{(x-y/2)}}\sigma) - i\Phi([\rho];{\bf{(x+y/2)}}\sigma)]
\]
\[
exp(U_{FS}[\{\rho({\bf{y^{'}}}\sigma^{'})
-\delta({\bf{y^{'} - (x-y/2)}})\delta_{\sigma^{'}, \sigma} \}]
-U_{FS}[\{\rho({\bf{y^{'}}}\sigma^{'})
-\delta({\bf{y^{'} - (x+y/2)}})\delta_{\sigma^{'}, \sigma} \}])
\]
\begin{equation}
\frac{\Theta([\{\rho({\bf{y^{'}}}\sigma^{'})
-\delta({\bf{y^{'} - (x+y/2)}})\delta_{\sigma^{'}, \sigma} \}]) }
{\Theta([\{\rho({\bf{y^{'}}}\sigma^{'})
-\delta({\bf{y^{'} - (x-y/2)}})\delta_{\sigma^{'}, \sigma} \}])}
 =  - n_{0}({\bf{y}}\sigma)
\label{eqnforphi}
\end{equation}
 The terms involving the Heaviside are indeterminate unless we expand
 around
\begin{equation}
\rho({\bf{y}}\sigma) = \delta^{d}({\bf{0}}) + \delta \rho({\bf{y}}\sigma)
\end{equation}
and,
\begin{equation}
\langle \mbox{  } \rho({\bf{y}}\sigma) \mbox{  } \rangle
 = \delta^{d}({\bf{0}}) = \rho_{\sigma}
\end{equation}

\subsection{Point Splitting or no Point Splitting ?}

 The attempts made here are partly based on the work of Ligouri and
  Mintchev on Generalised statistics\cite{Lig} and work of
  Goldin et. al.\cite{Sharp} and the series by
  Reed and Simon\cite{Reed}
 Here I shall attempt to provide a framework within which questions
 such as the existence of the canonical conjugate of the
 fermi-density distribution may be addressed.
 For reasons of clarity we shall not insist on utmost generality.
 The philosophy being that the quest for utmost genarality should
 not cloud the underlying basic principles.
 We start off with some preliminaries. Let $ {\mathcal{H}} $ be an
 infinite dimensional separable Hilbert Space. We know from
 textbooks that such a space possesses a countable orthonormal
 basis $ {\mathcal{B}} = \{ w_{i}; i \in {\mathcal{Z}} \} $ .
 Here, $ {\mathcal{Z}} $ is the set of all integers.
 Thus$ {\mathcal{H}} $ = Set of all linear combinations of vectors chosen
 from $ {\mathcal{B}} $. We construct the tensor product of two such
 spaces
\[
 {\mathcal{H}}^{{\small{\bigotimes}} 2}
 = {\mathcal{H}} {\bigotimes} {\mathcal{H}}
\]
 This is defined to be the dual space of the space of all bilinear
 forms on the direct sum. In plain English this means something like this.
 Let $ f \in {\mathcal{H}} $ and $ g \in {\mathcal{H}} $ define the
 object $ f {\small{\bigotimes}} g $
 to be that object which acts as shown below.
 Let $ < v, w > $ be an element of the Cartesian product
 $ {\mathcal{H}} \times {\mathcal{H}} $.
\[
 f {\small{\bigotimes}} g < v, w > = (f, v) (g, w)
\]
Here, $ (f, v) $ stands for the inner product of $ f $ and $ v $.
Define also the inner product of two $ f {\small{\bigotimes}} g $ and
  $ f^{'} {\small{\bigotimes}} g^{'} $
\[
(f {\small{\bigotimes}} g,  f^{'} {\small{\bigotimes}}
 g^{'}) = (f,  f^{'} ) (g, g^{'})
\]
 Construct the space of all finite linear combinations of
 objects such as $ f {\small{\bigotimes}} g $ with different choices for
 $ f $ and $ g $. Lump them all into a set. You get a vector space.
 It is still not the vector space
 $ {\mathcal{H}}^{{\small{\bigotimes 2}}} $.
 Because the space of all finite linear combinations of
 objects such as $ f {\small{\bigotimes}} g $ is not complete. Not every
 Cauchy sequence converges. Complete the space by appending the limit points
 of all Cauchy sequences from the space of all finite linear combinations
 of vectors of the type $ f {\small{\bigotimes}} g $.
 This complete space is the
 Hilbert space $ {\mathcal{H}}^{{\small{\bigotimes 2}}} $. Similarly one can
 construct $  {\mathcal{H}}^{{\small{\bigotimes}} n} $ for n = 0, 1, 2, 3, ...
 Where we have set $ {\mathcal{H}}^{0}  = {\mathcal{C}} $
 the set of complex numbers. Define the Fock Space over
 $ {\mathcal{H}} $ as
\[
{\mathcal{F}}({\mathcal{H}}) = {\bigoplus}_{n=0}^{\infty}
 {\mathcal{H}}^{{\small{\bigotimes}} n}
\]
Physically, each element of it is an ordered collection of wavefunctions
 with different number of particles
\[
 (\varphi_{0}, \varphi_{1}(x_{1}), \varphi_{2}(x_{1}, x_{2}),
 ...,\varphi_{n}, ... )
\]
 is a typical element of $ {\mathcal{F}}({\mathcal{H}}) $. This is the
 Hilbert Space which we shall be working with. Let $ {\mathcal{D}}^{n} $
 be the space of all decomposable vectors.
\[
{\mathcal{D}}^{n} = \{ f_{1} {\small{\bigotimes}}...{\small{\bigotimes}} f_{n}
; f_{i} \in {\mathcal{H}} \}
\]
For each $ f \in {\mathcal{H}} $ define
\[
b(f): {\mathcal{D}}^{n} \rightarrow {\mathcal{D}}^{n-1}, n \geq 1
\]
\[
b^{*}(f): {\mathcal{D}}^{n} \rightarrow {\mathcal{D}}^{n+1}, n \geq 0
\]
defined by
\[
b(f) \mbox{ } f_{1} {\small{\bigotimes}} ...  {\small{\bigotimes}} f_{n}
 = \sqrt{n} (f, f_{1}) f_{2} {\small{\bigotimes}} ...{\small{\bigotimes}}
 f_{n}
\]

\[
b^{*}(f) \mbox{ } f_{1} {\small{\bigotimes}} ...{\small{\bigotimes}} f_{n}
 = \sqrt{n+1}
 f \bigotimes  f_{1} {\small{\bigotimes}} f_{2}
{\small{\bigotimes}} ...  {\small{\bigotimes}} f_{n}
\]
We also define $ b(f) {\mathcal{H}}^{0}  = 0 $. By linearity we can extend
 the definitions to the space of all finite linear combinations
 of elements of $ {\mathcal{D}}^{n} $ namely $ {\mathcal{L}}
 ({\mathcal{D}}^{n}) $.
 For any $ \varphi \in {\mathcal{L}}({\mathcal{D}}^{n}) $ and
$ \psi \in {\mathcal{L}}({\mathcal{D}}^{n+1}) $
\[
\parallel b(f) \varphi \parallel \leq \sqrt{n} \parallel f \parallel
 \parallel \varphi \parallel
\]
\[
\parallel b^{*}(f) \varphi \parallel \leq \sqrt{n+1} \parallel f \parallel
 \parallel \varphi \parallel
\]
\[
(\psi,  b^{*}(f)\varphi) = ( b(f)\psi, \varphi)
\]
So long as $ \parallel f \parallel < \infty $, $  b(f) $ and $  b^{*}(f)  $
 are bounded operators. An operator $ {\mathcal{O}} $ is said to be bounded
 if
\[
 sup_{\parallel \varphi \parallel = 1}
\mbox{ }
 \parallel {\mathcal{O}} {\mathcal{\varphi}} \parallel \mbox{ }
 < \mbox{ } \infty
\]
 $ {\mathcal{O}} $ is unbounded otherwise. The norm of a bounded operator is
 defined as
\[
\parallel {\mathcal{O}} \parallel =  sup_{\parallel \varphi \parallel = 1}
\mbox{ }
 \parallel {\mathcal{O}} {\mathcal{\varphi}} \parallel \mbox{ }
\]
In order to describe fermions, it is necessary to
 construct orthogonal projectors on $ {\mathcal{F}}({\mathcal{H}}) $.
 In what follows $ c(f) $ will denote a fermi annhilation operator.
 $ c^{*}(f) $ will denote a fermi creation operator. Physically, and naively
 speaking, these are the fermi operators in "momentum space" $ c_{\bf{k}} $
 and $ c^{*}_{\bf{k}} $.
 First define $ P_{-} $ to be the projection operator that projects
 out only the antisymmetric parts of many body wavefunctions.
 For example,
\[
P_{-} f_{1} {\small{\bigotimes}} f_{2} = \frac{1}{2}
 (f_{1} {\small{\bigotimes}} f_{2} - f_{2} {\small{\bigotimes}} f_{1})
\]
We now have
\[
c(f) = P_{-}b(f)P_{-}
\]
\[
c^{*}(f) = P_{-}b^{*}(f)P_{-}
\]
Let us take a more complicated example. Let us find out how
 $ c^{*}(f) c(g) $ acts on a vector
 $ v =  f_{1} {\small{\bigotimes}} f_{2} $.
\[
c^{*}(f) c(g) = P_{-}b^{*}(f)P_{-}P_{-}b(g)P_{-}
\]
\[
c^{*}(f) c(g) = P_{-}b^{*}(f)P_{-}b(g)P_{-}
\]
\[
c^{*}(f) c(g) v = P_{-}b^{*}(f)P_{-}b(g)P_{-} v
\]
\[
P_{-} v = \frac{1}{2 !}( f_{1} {\small{\bigotimes}} f_{2}
 - f_{2} {\small{\bigotimes}} f_{1} )
\]

\[
b(g)P_{-} v = \frac{1}{2 !}{\sqrt{2}}( (g, f_{1}) f_{2} -
 (g, f_{2}) f_{1} )
\]
\[
P_{-}b(g)P_{-} v = \frac{1}{2 !}{\sqrt{2}}( (g, f_{1}) f_{2} -
 (g, f_{2}) f_{1} )
\]
\[
b^{*}(f)P_{-}b(g)P_{-} v = \frac{1}{2 !}{\sqrt{2}}^{2}
( (g, f_{1}) f {\small{\bigotimes}} f_{2} -
 (g, f_{2}) f {\small{\bigotimes}} f_{1} )
\]
\[
c^{*}(f) \mbox{ } c(g) v = (\frac{1}{2 !})^{2}{\sqrt{2}}^{2}
( (g, f_{1})[ f {\small{\bigotimes}}f_{2} - f_{2} {\small{\bigotimes}} f]
 -
 (g, f_{2}) [ f {\small{\bigotimes}} f_{1} - f_{1} {\small{\bigotimes}} f ] )
\]
 Having had a feel for how the fermi operators behave, we are now
 equipped to pose some more pertinent questions.
 Choose a basis
\[
{\mathcal{B}} = \{ w_{i}; i \in {\mathcal{Z}} \}
\]
                
\subsection{Definition of the Fermi Density Distribution}

 Here we would like to capture the notion of the fermi density operator.
 Physicists call it $ \rho(x) = \psi^{*}(x) \psi(x) $. Multiplication
 of two fermi fields at the same point is a tricky business and we would
 like to make more sense out of it. For this we have to set our single
 particle Hilbert Space:
\[
 {\mathcal{H}} = L_{p}({\mathcal{R}}^{3}) {\bigotimes} {\mathcal{W}}
\]
Here, $ L_{p}({\mathcal{R}}^{3}) $ is the space of all periodic functions
 with period $ L $ in each space direction. That is if
 $ u \in  L_{p}({\mathcal{R}}^{3}) $ then
\[
u(x_{1} + L, x_{2}, x_{3}) = u(x_{1}, x_{2}, x_{3})
\]
\[
u(x_{1}, x_{2} + L, x_{3}) = u(x_{1}, x_{2}, x_{3})
\]
\[
u(x_{1}, x_{2}, x_{3} + L) = u(x_{1}, x_{2}, x_{3})
\]

$ {\mathcal{W}} $ is the spin space spanned by two vectors.
 An orthonormal basis for $ {\mathcal{W}} $
\[
\{ \xi_{\uparrow}, \xi_{\downarrow} \}
\]
A typical element of $ {\mathcal{H}} $ is given by
 $ f({\bf{x}}) \bigotimes \xi_{\downarrow} $. A basis for $ {\mathcal{H}} $
 is given by
\[
{\mathcal{B}} =
 \{ \sqrt{ \frac{1}{L^{3}} } exp(i{\bf{q_{n}.x}}) \bigotimes
 \xi_{s}
  ;
 {\bf{n}} = (n_{1}, n_{2}, n_{3}) \in {\mathcal{Z}}^{3},
  s \in \{ \uparrow, \downarrow \} ;
\]
\[
 { \bf{q_{n}} } =
 (\frac{2 \pi n_{1}}{L}, \frac{2 \pi n_{2}}{L}, \frac{2 \pi n_{3}}{L})
 \}
\]
We move on to the definition of the fermi-density distribution.
 The Hilbert Space $ {\mathcal{H}}^{\bigotimes n} $ is the space of
 all n-particle wavefunctions with no symmetry restrictions.
 From this we may construct orthogonal subspaces
\[
{\mathcal{H}}_{+}^{\bigotimes n} = P_{+} {\mathcal{H}}^{\bigotimes n}
\]
\[
{\mathcal{H}}_{-}^{\bigotimes n} = P_{-} {\mathcal{H}}^{\bigotimes n}
\]
 Tensors from $ {\mathcal{H}}_{+}^{\bigotimes n} $ are orthogonal
 to tensors from $ {\mathcal{H}}_{-}^{\bigotimes n} $. The only exceptions are
 when $ n = 0 $ or $ n = 1 $.
\[
{\mathcal{H}}_{+}^{\bigotimes 0} = {\mathcal{H}}_{+}^{\bigotimes 0} =
 {\mathcal{C}}
\]
\[
{\mathcal{H}}_{+}^{\bigotimes 1} = {\mathcal{H}}_{+}^{\bigotimes 1} =
 {\mathcal{H}}
\]
The space $ {\mathcal{H}}_{+}^{\bigotimes n} $ is the space of
 bosonic-wavefunctions and the space $ {\mathcal{H}}_{-}^{\bigotimes n} $
 is the space of fermionic wavefunctions. The definition of the fermi
 density distribution proceeds as follows. Let $ v $ be written as
\[
 v = \sum_{\sigma \in \{ \uparrow, \downarrow \} }a(\sigma)\xi_{\sigma}
\]
 The Fermi density distibution is an operator on the Fock Space,
 given a vector $ f \bigotimes v \in \mathcal{H} $ in the single particle
 Hilbert Space, and a tensor $ \varphi $ in the n-particle subspace of
 of $ \mathcal{F}(\mathcal{H}) $, there exists a corresponding operator
 $ \rho(f \bigotimes v) $ that acts as follows:
\[
[\rho(f \bigotimes v) \varphi ]_{n}
({\bf{x_{1}}}\sigma_{1}, {\bf{x_{2}}}\sigma_{2},
 .... , {\bf{x_{n}}}\sigma_{n})
  = 0
\]
  if $ \varphi \in  {\mathcal{H}}_{+}^{\bigotimes n} $ and
\[
[\rho(f \bigotimes v) \varphi]_{n}
 ({\bf{x_{1}}}\sigma_{1}, {\bf{x_{2}}}\sigma_{2},
 .... , {\bf{x_{n}}}\sigma_{n})
 = \sum_{i=1}^{n} f({\bf{x_{i}}}) a(\sigma_{i})
 \varphi_{n}({\bf{x_{1}}}\sigma_{1}, {\bf{x_{2}}}\sigma_{2},
 .... , {\bf{x_{n}}}\sigma_{n})
\]
when $ \varphi \in  {\mathcal{H}}_{-}^{\bigotimes n} $.
The physical meaning of this abstract operator will become clear
 in the next subsection.

\subsection{Definition of the Canonical Conjugate of the Fermi Density}

 We introduce some notation.
 Let
$ g = \sqrt{ \frac{1}{L^{3}} }exp(i{ \bf{k_{m}.x} } ) \bigotimes \xi_{r} $
\[
\psi({ { \bf{k_{m}} } r}) = c(g)
\]
\[
\rho({ { \bf{k_{m}} } r}) = \rho(g)
\]
This $ \rho({ { \bf{k_{m}} } r})  $ is nothing but the density operator
 in momentum space, familiar to Physicists
\[
 \rho({{\bf{k_{m}}}r}) =
 \sum_{ {\bf{q_{n}}} }c^{\dagger}_{ {\bf{q_{n}+k_{m}}}r }c_{ {\bf{q_{n}}}r }
\]
 We want to define the canonical conjugate of the density operator
 as an operator that maps the Fock space(or a subset thereof)
 on to itself.
 If $ \varphi \in {\mathcal{H}}^{0} = {\mathcal{C}} $ then
\[
X_{ { \bf{q_{m}} } } \varphi = 0
\]
 Let $ \varphi \in  {\mathcal{H}}_{+}^{n} $, $ n = 2, 3, ... $ then
\[
X_{ { \bf{q_{m}} } } \varphi = 0
\]
 The important cases are when
 $ \varphi \in  {\mathcal{H}}_{-}^{n} $, $ n = 2, 3, ... $ or
 if $ \varphi \in  {\mathcal{H}} $. In such a case, we set
 $ n = N_{s}^{0} \neq 0 $ for $ s \in \{ \downarrow, \uparrow \} $.
 Let us introduce some more notation.
 $ N_{s} = \rho_{ { \bf{q = 0} } s } $ is the number operator to be
 distinguished from the c-number $ N_{s}^{0} $. The eigenvalue of
 $ N_{s} $ is $ N_{s}^{0} $ when it acts on a state such as
 $ \varphi \in {\mathcal{H}}_{-}^{n} $. Some more notation.
\[
  \delta \mbox{ } \psi({ { \bf{k_{m}} } s })
  =  \psi({ { \bf{k_{m}} } s }) - { \sqrt{N_{s}^{0}} }
\delta_{ { \bf{k_{m}, 0} } }
\]
 and
\[
 \delta \mbox{ } \rho({ { \bf{k_{m}} } s })
 = \rho({ { \bf{k_{m}} } s }) - N_{s}^{0} \delta_{ { \bf{k_{m}, 0} } }
\]
The Canonical Conjugate of the density distribution in real space
 denoted by $ \Pi_{s}({\bf{x}}) $ is defined as
 follows.
\[
\Pi_{s}(x_{1}+L, x_{2}, x_{3}) = \Pi_{s}(x_{1}, x_{2}, x_{3})
\]
\[
\Pi_{s}(x_{1}, x_{2}+L, x_{3}) = \Pi_{s}(x_{1}, x_{2}, x_{3})
\]
\[
\Pi_{s}(x_{1}, x_{2}, x_{3}+L) = \Pi_{s}(x_{1}, x_{2}, x_{3})
\]
The definition is as follows.
\[
\Pi_{s}({\bf{x}}) = \sum_{ {\bf{q_{m}}}}
 exp(i{\bf{q_{m}.x}})X_{ {\bf{q_{m}}}s }
\]

\[
 X_{ {\bf{q_{m}}}s } =
 i \mbox{ } {\large{ln}} [ ( 1 +
\frac{1}{ { \sqrt{N_{s}^{0}} } }
 \sum_{ { \bf{k_{n}} } } \delta \psi({ \bf{k_{n}} } s)
 T_{- {\bf{k_{n}}} }({\bf{q_{m}}}))( 1 + \sum_{ { \bf{k_{n}} } }
\]
                                                       
\begin{equation}
\frac{1}{ N_{s}^{0} } \delta \rho({ \bf{k_{n}} } s)
 T_{{\bf{k_{n}}} }({\bf{q_{m}}}) )^{-\frac{1}{2}}
exp(-i\sum_{ { \bf{k_{n}} } }\phi([\rho]; {\bf{k_{n}}} s)
T_{{\bf{k_{n}}} }({\bf{q_{m}}})  ) ] \delta_{ {\bf{q_{m}}}, 0}
\label{result}
\end{equation}
\[
\Phi([\rho]; {\bf{x}}s) = \sum_{ {\bf{k_{n}}} }
\phi([\rho];{\bf{k_{n}}} s) exp(-i{\bf{k_{n}.x}})
\]
\[
T_{{\bf{k_{n}}} }({\bf{q_{m}}}) = exp({\bf{k_{n}.\nabla_{q_{m}}}})
\]
 The translation operator translates the $ {\bf{q_{m}}} $
 in the Kronecker delta that appears in the extreme right by
 $ {\bf{k_{n}}} $
 and $ \Phi([\rho]; {\bf{x}}s) $ satisfies a recursion explained in detail
 in the previous manuscript. The logarithm is to be interpreted as
 an expansion around the leading term  which is either $ N_{s}^{0} $ or
 $ { \sqrt{N_{s}^{0}} } $. The question of existence of $ X_{ {\bf{q_{m}}}s } $
 now reduces to demonstrating that this operator (possibly unbounded)
 maps its domain of definition(densely defined in Fock space)
  on to the Fock space. Defining the limit of the series expansion
 is likely to be the major bottleneck in demonstrating the existence
 of $ X_{ {\bf{q_{m}}}s } $. That this is the canonical conjugate of the
 density operator is not at all obvious from the above definition. A
 rigorous proof of that is also likely to be difficult. Considering
 that we arrived at this formula by first postulating the existence of
 $ \Pi_{s}({\bf{x}}) $, it is probably safe to just say
 " it is clear that "  $ \Pi_{s}({\bf{x}}) $, is in fact the canonical
 conjugate of $ \rho $. The way in which the above formula can be
 deduced may be motivated as follows:
\[
 \psi({\bf{x}}\sigma) =
 \frac{ 1 }{ V^{\frac{1}{2}} }
\sum_{ {\bf{k}} }exp(i{\bf{k.x}})\psi({\bf{k}}\sigma)
\]
\[
 = exp(-i \sum_{{\bf{q}}} exp(i{\bf{q.x}}) X_{{\bf{q}}\sigma} )
exp(i \sum_{{\bf{q}}} exp(-i{\bf{q.x}}) \phi( [\rho]; {\bf{q}}\sigma ) )
\]                                         
\begin{equation}
(n_{\sigma} + \frac{1}{V} \sum_{{\bf{q}}\neq 0}
 \rho_{{\bf{q}}\sigma} exp(-i{\bf{q.x}}) )^{\frac{1}{2}}
\label{FF}
\end{equation}
In the above Eq.(~\ref{FF}) ONLY on the right side make the replacements,
\begin{equation}
exp(i{\bf{q.x}}) \rightarrow T_{{\bf{-q}}}({\bf{k}})
\end{equation}
\begin{equation}
exp(-i{\bf{q.x}}) \rightarrow T_{{\bf{q}}}({\bf{k}})
\end{equation}
where
\begin{equation}
T_{{\bf{q}}}({\bf{k}}) = exp({\bf{q}}.\nabla_{{\bf{k}}})
\end{equation}
and append a $ \delta_{ {\bf{k, 0}} } $ on the extreme right.
 Also on the LEFT side of Eqn.(~\ref{FF})
 make the replacement $ \psi({\bf{x}}\sigma) $ by
 $ \psi({\bf{k}}\sigma) $. And you get a formula for
 $  \psi({\bf{k}}\sigma) $.
 In order to get a formula for $ X_{ {\bf{q_{m}} } s } $
 we have to invert the relation and obtain,
\[
\Pi({\bf{x}}s) =
i \mbox{ } ln[  ( \sqrt{N_{s}^{0}} +
\sum_{ {\bf{k_{n}}}}
exp(i{\bf{k_{n}.x}})\delta\psi({\bf{k_{n}}}s) )
\]
\begin{equation}
exp(-i \sum_{{\bf{k_{n}}}} exp(-i{\bf{k_{n}.x}})
 \phi( [\rho]; {\bf{k_{n}}}s ) )(N_{s}^{0} + \sum_{ {\bf{k_{n}}} }
\delta\rho_{{\bf{k_{n}}}s} exp(-i{\bf{k_{n}.x}}) )^{-\frac{1}{2}} ]
\label{canon}
\end{equation}
Make the replacements on the right side of Eqn.(~\ref{canon})
\begin{equation}
exp(i{\bf{k_{n}.x}}) \rightarrow T_{{\bf{-k_{n}}}}({\bf{q_{m}}})
\end{equation}
\begin{equation}
exp(-i{\bf{k_{n}.x}}) \rightarrow T_{{\bf{k_{n}}}}({\bf{q_{m}}})
\end{equation}
where
and append a $ \delta_{ {\bf{q_{m}, 0}} } $ on the extreme right.
Also on the LEFT side of Eqn.(~\ref{canon}) replace
$ \Pi({\bf{x}}s) $ by $ X_{ {\bf{q_{m}}}s } $. This results in formula
 given in Eqn. (~\ref{result}).

\section{Expression in terms of Fermi Sea Displacemnts}

The claim is that the following exact relation holds,
\[
c^{\dagger}_{ {\bf{k+q/2}} }c_{ {\bf{k-q/2}} }
 = n_{F}({\bf{k}})\frac{N}{\langle N \rangle}\delta_{ {\bf{q, 0}} }  +
 (\sqrt{ \frac{N}{\langle N \rangle} })
[\Lambda_{ {\bf{k}} }( {\bf{q}} )
\theta( -{\bf{k.q}} )
 a_{ {\bf{k}} }({\bf{-q}}) + \Lambda^{*}_{ {\bf{k}} }( {\bf{-q}} )
\theta( {\bf{k.q}} )
a^{\dagger}_{ {\bf{k}} }({\bf{q}}) ]
\]
\begin{equation}
+ \sum_{ { \bf{k_{1}, k_{2}} } }\sum_{ { \bf{q_{1}, q_{2}} } }
\Gamma^{{ \bf{q_{1}, q_{2}} } }
_{ { \bf{k_{1}, k_{2}} } }
({ \bf{k, q} })a^{\dagger}_{ {\bf{k}}_{1} }( {\bf{q}}_{1} )
a_{ {\bf{k}}_{2} }( {\bf{q}}_{2} )
\label{SEA}
\end{equation}
where,
\begin{equation}
\{c_{ {\bf{k}} }, c_{ {\bf{k}}^{'} } \} = 0
\end{equation}
and
\begin{equation}
\{c_{ {\bf{k}} }, c^{\dagger}_{ {\bf{k}}^{'} } \} =
 \delta_{ {\bf{k}}, {\bf{k}}^{'} }
\end{equation}
and,
\begin{equation}
[ a_{ {\bf{k}} }({\bf{q}}), a_{ {\bf{k}}^{'} }({\bf{q}}^{'}) ]
 = 0
\end{equation}
\begin{equation}
[ a_{ {\bf{k}} }({\bf{q}}), a^{\dagger}_{ {\bf{k}}^{'} }({\bf{q}}^{'}) ]
 = \delta_{ {\bf{k}}, {\bf{k}}^{'} }\delta_{ {\bf{q}}, {\bf{q}}^{'} }
\end{equation}
and,
\begin{equation}
\Lambda_{ {\bf{k}} }({\bf{q}})
 = \sqrt{ n_{F}({\bf{k+q/2}})(1 - n_{F}({\bf{k-q/2}})) }
\end{equation}
and
\begin{equation}
N = \sum_{ {\bf{k}} }c^{\dagger}_{ {\bf{k}} }c_{ {\bf{k}} }
\end{equation}
\begin{equation}
\langle N \rangle = \sum_{ {\bf{k}} }n_{F}({\bf{k}})
\end{equation}
and $ n_{F}({\bf{k}}) = \theta(k_{f} - |{\bf{k}}|) $
The $ \theta(-{\bf{k.q}}) $ is a reminder that
 $ \Lambda_{ {\bf{k}} }({\bf{q}}) $ is nonzero only when
 $ {\bf{k.q}} \leq 0 $.
Also,
\begin{equation}
[N, a_{ {\bf{k}} }({\bf{q}}) ] = 0
\end{equation}
\begin{equation}
[c_{ {\bf{p}} }, a_{ {\bf{k}} }({\bf{q}}) ] \mbox{    } =
 something  \mbox{    } complicated
\end{equation}
and the kinetic energy operator is,
\[
  K = \sum_{ {\bf{k}} }\epsilon_{ {\bf{k}} }
c^{\dagger}_{ {\bf{k}} }c_{ {\bf{k}} } 
\]
in terms of the Bose fields, it is postulated to be,
\begin{equation}
K = E_{0} + \sum_{ {\bf{k}}, {\bf{q}} } \omega_{ {\bf{k}} }( {\bf{q}} )
 a^{\dagger}_{ {\bf{k}} }( {\bf{q}} )
 a_{ {\bf{k}} }( {\bf{q}} )
\label{Kin}
\end{equation}
\[
\omega_{ {\bf{k}} }( {\bf{q}} ) = \Lambda_{{\bf{k}}}(-{\bf{q}})
\frac{ {\bf{k.q}} }{m}
\]
and $ E_{0} = \sum_{ {\bf{k}} }\epsilon_{ {\bf{k}} }n_{F}({\bf{k}}) $.
Also, the filled Fermi sea is identified with the Bose vacuum.
 $ |FS> = |0> $.
The fact that this is true will become clearer later. At this point
 it is sufficient to note that 
$ \omega_{ {\bf{k}} }( {\bf{q}} )  \geq 0 $ for all $ {\bf{k}} $ and 
 $ {\bf{q}} $. 
We have to make sure that for the right choice of $ \Gamma $,
\begin{equation}
  \langle \rho_{ {\bf{q_{1}}} }(t_{1})\rho_{ {\bf{q_{2}}} }(t_{2})
\rho_{ {\bf{-q_{1}-q_{2}}} }(t_{3}) \rangle \mbox{  }\neq  \mbox{  } 0
\end{equation}
 The terms linear in the Fermi sea displacements in Eq.(~\ref{SEA}) are
 chosen in order to ensure that the correct dynamical four-point functions
 are recovered.  
 Now for the quadratic terms. The best way to derive
 formulae for them is to compute the exact dynamical six point
 function and set it equal to the expression got from the free theory.
\[
 \langle c^{\dagger}_{ {\bf{k+q/2}} }(t)c_{ {\bf{k-q/2}} }(t)
c^{\dagger}_{ {\bf{k^{'}+q^{'}/2}} }(t^{'})c_{ {\bf{k^{'}-q^{'}/2}} }(t^{'})
c^{\dagger}_{ {\bf{k^{''}+q^{''}/2}} }(t^{''})
c_{ {\bf{k^{''}-q^{''}/2}} }(t^{''}) \rangle
 = I({\bf{k,q}},t;{\bf{k^{'},q^{'}}},t^{'};{\bf{k^{''},q^{''}}},t^{''})
\]
From the free theory we have
\[
 I({\bf{k,q}},t;{\bf{k^{'},q^{'}}},t^{'};{\bf{k^{''},q^{''}}},t^{''})
= exp(i\frac{ {\bf{k.q}} }{m}t)exp(i\frac{ {\bf{k^{'}.q^{'}}} }{m}t^{'})
exp(i\frac{ {\bf{k^{''}.q^{''}}} }{m}t^{''})
\]
\[
[\mbox{   }
( 1 - n_{F}({\bf{k-q/2}}) ) ( 1 - n_{F}({\bf{k^{'}-q^{'}/2}}) )
n_{F}({\bf{k+q/2}})
\delta_{ {\bf{k+q/2}}, {\bf{k^{''}-q^{''}/2}} }
\delta_{ {\bf{k-q/2}}, {\bf{k^{'}+q^{'}/2}} }
\delta_{ {\bf{k^{'}-q^{'}/2}}, {\bf{k^{''}+q^{''}/2}} }
\]
\begin{equation}
- \delta_{ {\bf{k-q/2}}, {\bf{k^{''}+q^{''}/2}} }
\delta_{ {\bf{k+q/2}}, {\bf{k^{'}-q^{'}/2}} }
\delta_{ {\bf{k^{'}+q^{'}/2}}, {\bf{k^{''}-q^{''}/2}} }
(1 - n_{F}({\bf{k-q/2}}) )n_{F}({\bf{k^{'}+q^{'}/2}})
n_{F}({\bf{k+q/2}}) \mbox{   }]
\label{Free}
\end{equation}
Also the six-point function may be evaluated in terms of the
 Bose fields(summation over superfluous indices is implied),
\[
I =
\langle
\Lambda_{ {\bf{k}} }({\bf{q}})a_{ {\bf{k}} }(-{\bf{q}})
exp(-i\omega_{ {\bf{k}} }(-{\bf{q}})t)
\Gamma_{ {\bf{k_{1}, k_{2}}} }^{ {\bf{q_{1}, q_{2}}} }
({\bf{k}}^{'}, {\bf{q}}^{'})a^{\dagger}_{ {\bf{k_{1}}} }({\bf{q_{1}}})
a_{ {\bf{k_{2}}} }({\bf{q_{2}}})
\]
\begin{equation}
exp(i\omega_{ {\bf{k_{1}}} }({\bf{q_{1}}})t^{'})
exp(-i\omega_{ {\bf{k_{2}}} }({\bf{q_{2}}})t^{'})
\Lambda_{ {\bf{k^{''}}} }(-{\bf{q^{''}}})
a^{\dagger}_{ {\bf{k^{''}}} }({\bf{q}}^{''})
exp(i\omega_{ {\bf{k^{''}}} }( {\bf{q}}^{''})t^{''}) \rangle
\label{Bose}
\end{equation}

The reduced form of the six-point function may be written as,
\[
I = \Lambda_{ {\bf{k}} }({\bf{q}})
exp(-i\omega_{ {\bf{k}} }(-{\bf{q}})t)
\Gamma_{ {\bf{k}}_{1}, {\bf{k}}_{2} }^{ {\bf{q}}_{1}, {\bf{q}}_{2} }
({\bf{k}}^{'}, {\bf{q}}^{'})
exp(i\omega_{ {\bf{k}}_{1} }({\bf{q}}_{1})t^{'})
exp(-i\omega_{ {\bf{k}}_{2} }({\bf{q}}_{2})t^{'})
\Lambda_{ {\bf{k}}^{''} }(-{\bf{q}}^{''})
exp(i\omega_{ {\bf{k}}^{''} }({\bf{q}}^{''})t^{''})
\]
\[
\langle a_{ {\bf{k}} }(-{\bf{q}})
a^{\dagger}_{ {\bf{k}}_{1} }({\bf{q}}_{1})
a_{ {\bf{k}}_{2} }({\bf{q}}_{2})
a^{\dagger}_{ {\bf{k}}^{''} }({\bf{q}}^{''}) \rangle
\]
This may be simplified to
\[
I = exp(-i\mbox{ }\omega_{ {\bf{k}} }(-{\bf{q}})\mbox{ }t)
exp(i\mbox{ }\omega_{ {\bf{k}} }(-{\bf{q}})\mbox{ }t^{'})
exp(-i\mbox{ }\omega_{ {\bf{k}}^{''} }({\bf{q}}^{''})\mbox{ }t^{'})
exp(i\mbox{ }\omega_{ {\bf{k}}^{''} }({\bf{q}}^{''})\mbox{ }t^{''})
\Lambda_{ {\bf{k}} }( {\bf{q}} )
\]
\begin{equation}
\Lambda_{ {\bf{k}}^{''} }( -{\bf{q}}^{''} )
\Gamma_{ {\bf{k}}, {\bf{k}}^{''} }^{ -{\bf{q}}, {\bf{q}}^{''} }
( {\bf{k}}^{'}, {\bf{q}}^{'} )
\end{equation}
Further, since the commutation rules,
\begin{equation}
[ c^{\dagger}_{ {\bf{k+q/2}} }c_{ {\bf{k-q/2}} }, 
 c^{\dagger}_{ {\bf{k^{'}+q^{'}/2}} }c_{ {\bf{k^{'}-q^{'}/2}} } ] 
 = c^{\dagger}_{ {\bf{k+q/2}} }c_{ {\bf{k^{'}-q^{'}/2}} } 
\delta_{ {\bf{k-q/2}}, {\bf{k^{'}+q^{'}/2}} }
 - c^{\dagger}_{ {\bf{k^{'}+q^{'}/2}} }c_{ {\bf{k-q/2}} }
 \delta_{ {\bf{k+q/2}}, {\bf{k^{'}-q^{'}/2}} }
\end{equation}
This means, that the coefficents also have to satisfy,
\[
-\Gamma^{-q^{'}\mbox{ }q_{2}}_{k^{'}\mbox{ }k_{2}}(k,q)
\Lambda_{k^{'}}(q^{'})
 + \Gamma^{-q\mbox{ }q_{2}}_{k \mbox{ }k_{2}}(k^{'}q^{'})
\Lambda_{k}(q)
\]
\begin{equation}
= 
\Lambda_{k^{'}+q/2}(q+q^{'})\delta_{k-q/2, k^{'}+q^{'}/2}
\delta_{k_{2}, k^{'}+q/2}\delta_{q_{2}, -q-q^{'} }
 - \Lambda_{k^{'}-q/2}(q+q^{'})\delta_{k+q/2, k^{'}-q^{'}/2}
\delta_{k_{2}, k^{'}-q/2}\delta_{q_{2}, -q-q^{'} }
\label{Comm}
\end{equation}
First notice that $ \Lambda_{ {\bf{k}} }({\bf{q}}) $ is either zero or
 one.
Notice that Eq.(~\ref{Free}), Eq.(~\ref{Bose}) and
 Eq.(~\ref{Comm}) are such that 
 they suggest to us,
\begin{equation}
\Gamma_{ {\bf{k}}_{1}, {\bf{k}}_{2} }^{ {\bf{q}}_{1}, {\bf{q}}_{2} }
({\bf{k}}, {\bf{q}})
 = 0\mbox{    } 
 unless\mbox{    } \Lambda_{ {\bf{k}}_{1} }(-{\bf{q}}_{1}) = 1
 \mbox{    } and \mbox{    }
\Lambda_{ {\bf{k}}_{2} }(-{\bf{q}}_{2}) = 1
\end{equation}


\[
\Gamma_{ {\bf{k}}_{1}, {\bf{k}}_{2} }^{ {\bf{q}}_{1}, {\bf{q}}_{2} }
({\bf{k}}, {\bf{q}})
 =  \delta_{ {\bf{k}}_{1} - {\bf{q}}_{1}/2 , {\bf{k}}_{2} - {\bf{q}}_{2}/2}
 \delta_{ {\bf{k}}_{1} + {\bf{q}}_{1}/2 , {\bf{k}}+ {\bf{q}}/2 }
 \delta_{ {\bf{k}} - {\bf{q}}/2 , {\bf{k}}_{2} + {\bf{q}}_{2}/2 }
\]
\begin{equation}
 - \delta_{ {\bf{k}}_{1} + {\bf{q}}_{1}/2 , {\bf{k}}_{2} + {\bf{q}}_{2}/2}
 \delta_{ {\bf{k}}_{1} - {\bf{q}}_{1}/2 , {\bf{k}} - {\bf{q}}/2 }
 \delta_{ {\bf{k}} + {\bf{q}}/2 , {\bf{k}}_{2} - {\bf{q}}_{2}/2 }
 \mbox{       } for \mbox{       } 
\Lambda_{ {\bf{k_{1}}} }(-{\bf{q_{1}}})
\Lambda_{ {\bf{k_{2}}} }(-{\bf{q_{2}}}) = 1
\end{equation}
and
\begin{equation}
\Gamma_{ {\bf{k}}_{1}, {\bf{k}}_{2} }^{ {\bf{q}}_{1}, {\bf{q}}_{2} }
({\bf{k}}, {\bf{q}})
 = 0 \mbox{       } for \mbox{       }
\Lambda_{ {\bf{k_{1}}} }(-{\bf{q_{1}}})
\Lambda_{ {\bf{k_{2}}} }(-{\bf{q_{2}}}) = 0
\end{equation}
From this one may verify that the formula for the kinetic energy operator
 written down previously, Eq.(~\ref{Kin}) is correct.
Lastly, we must make sure that the following additional commutation rule holds.
\[
\Gamma_{ {\bf{k_{1}}}, {\bf{k_{3}}} }^{  {\bf{q_{1}}}, {\bf{q_{3}}} }
({\bf{k}}, {\bf{q}})\mbox{ }\Gamma_{ {\bf{k_{3}}}, {\bf{k_{2}}} }
^{  {\bf{q_{3}}}, {\bf{q_{2}}} }
({\bf{k}}^{'}, {\bf{q}}^{'})
 - \Gamma_{ {\bf{k_{3}}}, {\bf{k_{2}}} }^{  {\bf{q_{3}}}, {\bf{q_{2}}} }
({\bf{k}}, {\bf{q}})\mbox{ }\Gamma_{ {\bf{k_{1}}}, {\bf{k_{3}}} }
^{  {\bf{q_{1}}}, {\bf{q_{3}}} }
({\bf{k}}^{'}, {\bf{q}}^{'})
\]
\begin{equation}
 = \Gamma^{ {\bf{q_{1}}}, {\bf{q_{2}}} }_{ {\bf{k_{1}}}, {\bf{k_{2}}} }
({\bf{k}}^{'} + {\bf{q}}/2, {\bf{q+q^{'}}} )
\delta_{ {\bf{k-q/2}},  {\bf{k^{'}+q^{'}/2}} }
 - \Gamma^{ {\bf{q_{1}}}, {\bf{q_{2}}} }_{ {\bf{k_{1}}}, {\bf{k_{2}}} }
({\bf{k}} + {\bf{q}}^{'}/2, {\bf{q+q^{'}}} )
\delta_{ {\bf{k^{'}-q^{'}/2}},  {\bf{k+q/2}} }
\end{equation}
 Verification of the above rule completes the confirmation that
 the formula for $ c^{\dagger}_{ {\bf{k+q/2}} }c_{ {\bf{k-q/2}} } $
 in terms of the bose fields is exact. The algebra here goes through
 without a hitch expect in this respect, namely we have to assume that
 the $ \Lambda_{ {\bf{k}} }(-{\bf{q}}) = 1 $.

 To summarise let us write down the following:


***************************************************************************

 {\bf{THEOREM:}} 
\[
          c^{\dagger}_{ {\bf{k+q/2}} }c_{ {\bf{k-q/2}} }
 = n_{F}({\bf{k}})\frac{N}{\langle N \rangle }\delta_{ {\bf{q}}, {\bf{0}} }
 + 
(\sqrt{ \frac{N}{\langle N \rangle } })
[\Lambda_{ {\bf{k}} }({\bf{q}})a_{ {\bf{k}} }(-{\bf{q}})
 + \Lambda_{ {\bf{k}} }(-{\bf{q}})a^{\dagger}_{ {\bf{k}} }({\bf{q}}) ]
\]
\begin{equation}
 + \sum_{ {\bf{k}}_{1}, {\bf{k}}_{2} } 
 \sum_{ {\bf{q}}_{1}, {\bf{q}}_{2} }
  \Gamma_{ {\bf{k}}_{1}, {\bf{k}}_{2} }^{ {\bf{q}}_{1}, {\bf{q}}_{2} }
({\bf{k}}, {\bf{q}})a^{\dagger}_{ {\bf{k}}_{1} }({\bf{q}}_{1})
a_{ {\bf{k}}_{2} }({\bf{q}}_{2}) 
\end{equation}
 together with
\[
\Gamma_{ {\bf{k}}_{1}, {\bf{k}}_{2} }^{ {\bf{q}}_{1}, {\bf{q}}_{2} }
({\bf{k}}, {\bf{q}})
 = \Lambda_{ {\bf{k_{1}}} }(-{\bf{q_{1}}})
\Lambda_{ {\bf{k_{2}}} }(-{\bf{q_{2}}})
[ \delta_{ {\bf{k}}_{1} - {\bf{q}}_{1}/2 , {\bf{k}}_{2} - {\bf{q}}_{2}/2}
 \delta_{ {\bf{k}}_{1} + {\bf{q}}_{1}/2 , {\bf{k}}+ {\bf{q}}/2 }
 \delta_{ {\bf{k}} - {\bf{q}}/2 , {\bf{k}}_{2} + {\bf{q}}_{2}/2 }
\]
\begin{equation}
 - \delta_{ {\bf{k}}_{1} + {\bf{q}}_{1}/2 , {\bf{k}}_{2} + {\bf{q}}_{2}/2}
 \delta_{ {\bf{k}}_{1} - {\bf{q}}_{1}/2 , {\bf{k}} - {\bf{q}}/2 }
 \delta_{ {\bf{k}} + {\bf{q}}/2 , {\bf{k}}_{2} - {\bf{q}}_{2}/2 } ]
\end{equation}
 is an exact transformation of products of Fermi fields into Bose
 fields. Namely, all dynamical moments of 
$ c^{\dagger}_{ {\bf{k+q/2}} }c_{ {\bf{k-q/2}} } $ are recovered
 exactly, provided we identify the filled Fermi sea with the
 Bose vacuum.
\begin{equation}
a_{ {\bf{k}} }({\bf{q}}) |FS \rangle = 0
\end{equation}
and the above is also an operator identity, namely they satisfy proper mutual
 commutation rules.

******************************************************************************

 This may also be written in a simplified form as,
\[
          c^{\dagger}_{ {\bf{k+q/2}} }c_{ {\bf{k-q/2}} }
 = n_{F}({\bf{k}})\frac{N}{\langle N \rangle }\delta_{ {\bf{q}}, {\bf{0}} }
 + (\sqrt{ \frac{N}{\langle N \rangle } })
[\Lambda_{ {\bf{k}} }({\bf{q}})a_{ {\bf{k}} }(-{\bf{q}})
 + \Lambda_{ {\bf{k}} }(-{\bf{q}})a^{\dagger}_{ {\bf{k}} }({\bf{q}}) ]
\]
\[
 + 
 \sum_{ {\bf{q}}_{1} }\Lambda_{ {\bf{k+q/2 - q_{1}/2}} }(-{\bf{q_{1}}})
\Lambda_{ {\bf{k - q_{1}/2}} }({\bf{q-q_{1}}})
  a^{\dagger}_{ {\bf{k}} + {\bf{q}}/2 - {\bf{q}}_{1}/2 }({\bf{q}}_{1})
a_{ {\bf{k}} -  {\bf{q}}_{1}/2 }(-{\bf{q}} + {\bf{q}}_{1})
\]
\begin{equation}
 - \sum_{ {\bf{q}}_{1} }\Lambda_{ {\bf{k-q/2 + q_{1}/2}} }(-{\bf{q_{1}}})
\Lambda_{ {\bf{k + q_{1}/2}} }({\bf{q-q_{1}}})
  a^{\dagger}_{ {\bf{k}} - {\bf{q}}/2 + {\bf{q}}_{1}/2 }({\bf{q}}_{1})
a_{ {\bf{k}} + {\bf{q}}_{1}/2 }(-{\bf{q}} + {\bf{q}}_{1})
\end{equation}

***************************************************************************

 The $ \frac{N}{\langle N \rangle } $ is needed to ensure that,
 $ \sum_{ {\bf{k}} }c^{\dagger}_{ {\bf{k}} }c_{ {\bf{k}} } = N $
 (and not 
$ \sum_{ {\bf{k}} }c^{\dagger}_{ {\bf{k}} }c_{ {\bf{k}} } = \langle N \rangle $)
 The $ (\sqrt{ \frac{N}{\langle N \rangle } }) $ is needed to ensure that
 the commutation amongst the $ c^{\dagger}_{ {\bf{k+q/2}} }c_{ {\bf{k-q/2}} } $
 come out right.
 However, there are other finer points that need to be discussed.
 For example, do the thermodynamic expectation values also come out right ?
 That is, assuming that the temperature is finite, can we compute the
 dynamical density correlation function and show that it agrees with
 the free theory ? The finite temp. case is most unusual. It seems
 that for dealing with fermions at finite temperature, we have to assume
 that the Bose like-excitations are always at zero temperature with 
 zero chemical potential, only the coefficients in the
 expansion acquire finite temp. values. All this is very strange and
 on shaky ground. Thus, I shall relegate the finite temp.
 case to the last section where the skeptical reader may have a field day
 demolishing my ideas.

\subsection{The Fermi Field Operator}
 Also how about expressing the
 Fermi fields themselves in terms of the Fermi sea displacements ?
 For this one has to first construct a canonical conjugate of the
 density and then use it in the DPVA ansatz Eq.(~\ref{DPVA})
 \cite{Setlur} to compute the phase
 functional $ \Phi $. I have tried to construct a canonical conjugate
 of $ \rho $ ( any canonical conjugate suffices ) but it has not
 been successful. Just for future reference, it is useful to
 write down a formula for the total current operator.
\begin{equation}
{\bf{J}}_{tot} = \sum_{ {\bf{k}} }(\frac{ {\bf{k}} }{m})
c^{\dagger}_{ {\bf{k}} }
c_{ {\bf{k}} }
 = \sum_{ {\bf{k}}_{1}, {\bf{k}}_{2} }
 \sum_{ {\bf{q}}_{1}, {\bf{q}}_{2} }{\bf{\Gamma}}_{ {\bf{k}}_{1}, {\bf{k}}_{2} }
^{ {\bf{q}}_{1}, {\bf{q}}_{2} }a^{\dagger}_{ {\bf{k}}_{1} }
({\bf{q}}_{1})a_{ {\bf{k}}_{2} }({\bf{q}}_{2})
\end{equation}
\begin{equation}
{\bf{\Gamma}}_{ {\bf{k}}_{1}, {\bf{k}}_{2} }^{ {\bf{q}}_{1}, {\bf{q}}_{2} }
 = \Lambda_{ {\bf{k}}_{1} }(-{\bf{q}}_{1})
\Lambda_{ {\bf{k}}_{2} }(-{\bf{q}}_{2})
(\frac{ {\bf{q}}_{1} }{m})\delta_{ {\bf{k}}_{1}, {\bf{k}}_{2} }
 \delta_{ {\bf{q}}_{1}, {\bf{q}}_{2} }
\end{equation}
Therefore,
\begin{equation}
{\bf{J}}_{tot} = \sum_{ {\bf{k}} }(\frac{ {\bf{k}} }{m})
c^{\dagger}_{ {\bf{k}} }
c_{ {\bf{k}} }
 = \sum_{ {\bf{k}}_{1}, {\bf{q}}_{1} }
  \Lambda_{ {\bf{k}}_{1} }(-{\bf{q}}_{1})
(\frac{ {\bf{q}}_{1} }{m})a^{\dagger}_{ {\bf{k}}_{1} }
({\bf{q}}_{1})a_{ {\bf{k}}_{1} }({\bf{q}}_{1})
\end{equation}
**************************************************************************

Thus an alternative but less systematic approach seems to be in order.
 For this choose,
\begin{equation}
\psi({\bf{x}}) = exp( -i\mbox{ }U_{0}({\bf{x}}) )
exp( -i\mbox{ }U_{1}({\bf{x}}) )
exp( -i\mbox{ }U_{2}({\bf{x}}) )
( \rho({\bf{x}}) )^{\frac{1}{2}}
\end{equation}
here $ U_{0}({\bf{x}}) $ is a c-number and,
\begin{equation}
 U_{1}({\bf{x}}) = 
\sum_{ {\bf{k}}, {\bf{q}} }[a_{ {\bf{k}} }({\bf{q}})
f({\bf{k}}, {\bf{q}}; {\bf{x}}) + a^{\dagger}_{ {\bf{k}} }({\bf{q}})
f^{*}({\bf{k}}, {\bf{q}}; {\bf{x}})]
\end{equation}
\begin{equation}
U_{2}({\bf{x}}) =
\sum_{ {\bf{k}}_{1}, {\bf{q}}_{1} }
\sum_{ {\bf{k}}_{2}, {\bf{q}}_{2} }
U_{2}({\bf{k}}_{1}, {\bf{q}}_{1}; {\bf{k}}_{2}, {\bf{q}}_{2}; {\bf{x}})
a^{\dagger}_{ {\bf{k}}_{1} }({\bf{q}}_{1})
a_{ {\bf{k}}_{2} }({\bf{q}}_{2})
\end{equation}
Now let us evolve this using the kinetic energy operator,
\begin{equation}
\psi({\bf{x}},t) = 
 e^{i \mbox{ }t\mbox{ }K}\psi({\bf{x}})e^{-i \mbox{ }t\mbox{ }K}
 = exp(-i\mbox{ }U_{0}({\bf{x}}))
exp(-i\mbox{ }U_{1}({\bf{x}},t))
exp(-i\mbox{ }U_{2}({\bf{x}},t))( \rho({\bf{x}},t) )^{\frac{1}{2}}
\end{equation}
here,
\begin{equation}
 U_{1}({\bf{x}},t) = \sum_{ {\bf{k}}, {\bf{q}} }[a_{ {\bf{k}} }({\bf{q}})
e^{-i\mbox{ }t\mbox{ }\omega_{ {\bf{k}} }({\bf{q}})}
f({\bf{k}}, {\bf{q}}; {\bf{x}}) + a^{\dagger}_{ {\bf{k}} }({\bf{q}})
e^{i\mbox{ }t\mbox{ }\omega_{ {\bf{k}} }({\bf{q}})}
f^{*}({\bf{k}}, {\bf{q}}; {\bf{x}})]
\end{equation}

\begin{equation}
U_{2}({\bf{x}},t) =
\sum_{ {\bf{k}}_{1}, {\bf{q}}_{1} }
\sum_{ {\bf{k}}_{2}, {\bf{q}}_{2} }
U_{2}({\bf{k}}_{1}, {\bf{q}}_{1}; {\bf{k}}_{2}, {\bf{q}}_{2}; {\bf{x}})
a^{\dagger}_{ {\bf{k}}_{1} }({\bf{q}}_{1})
a_{ {\bf{k}}_{2} }({\bf{q}}_{2})
e^{i\mbox{ }t\mbox{ }
(\omega_{ {\bf{k}}_{1} }({\bf{q}}_{1})-
\omega_{ {\bf{k}}_{2} }({\bf{q}}_{2}))}
\end{equation}

 The basic philosophy it seems, involves observing that the commutation
 rules satisfied by number conserving objects such as
 $ c^{\dagger}_{ {\bf{k+q/2}} }c_{ {\bf{k-q/2}} } $ amongst themselves
 is independent of the underlying statistics. Thus one is lead to postulate
 convenient forms for such objects in terms of other Bose fields.
 But not all approaches are likely to be correct. Take for example, the
 straightforward substitution,
\begin{equation}
c^{\dagger}_{ {\bf{k+q/2}} }c_{ {\bf{k-q/2}} }
 = b^{\dagger}_{ {\bf{k+q/2}} }b_{ {\bf{k-q/2}} }
\end{equation}
 where  $ c_{ {\bf{k}} } $ are fermions and  $ b_{ {\bf{k}} } $ 
 are bosons. This no doubt makes the commutation rules come out right but
 is quite obviously wrong because all the correlation functions
 are wrongly represented.

\section{Luttinger and Fermi Liquids}
Consider an interaction of the type,
\begin{equation}
H_{I} = \sum_{ {\bf{q}} \neq 0 }\frac{ v_{ {\bf{q}} } }{2 \mbox{ }V}
\sum_{ {\bf{k}}, {\bf{k}}^{'} }
[\Lambda_{ {\bf{k}} }({\bf{q}})a_{ {\bf{k}} }(-{\bf{q}}) +
 \Lambda_{ {\bf{k}} }(-{\bf{q}})a^{\dagger}_{ {\bf{k}} }({\bf{q}})]
[\Lambda_{ {\bf{k}}^{'} }(-{\bf{q}})a_{ {\bf{k}}^{'} }({\bf{q}}) +
 \Lambda_{ {\bf{k}}^{'} }({\bf{q}})a^{\dagger}_{ {\bf{k}}^{'} }(-{\bf{q}})]
\end{equation}

%
\begin{equation}
i \mbox{ }\frac{\partial}{ \partial t }a^{t}_{ {\bf{k}} }({\bf{q}})
 = \omega_{ {\bf{k}} }({\bf{q}})a^{t}_{ {\bf{k}} }({\bf{q}})
 + (\frac{v_{ {\bf{q}} } }{V})
\Lambda_{ {\bf{k}} }(-{\bf{q}})\sum_{ {\bf{k}}^{'} }[
\Lambda_{ {\bf{k}}^{'} }(-{\bf{q}})a^{t}_{ {\bf{k}}^{'} }({\bf{q}})
+ \Lambda_{ {\bf{k}}^{'} }({\bf{q}})a^{\dagger t}_{ {\bf{k}}^{'} }(-{\bf{q}}) ]
\end{equation}
The equations of motion for the Bose propagators read as,
\[
(i\frac{\partial}{\partial t} - \omega_{ {\bf{k}} }({\bf{q}}))
\frac{ -i\langle T a^{t}_{ {\bf{k}} }({\bf{q}})
a^{\dagger}_{ {\bf{k}}^{'} }({\bf{q}}^{'}) \rangle }{\langle T 1 \rangle}
 = \delta_{ {\bf{k}}, {\bf{k}}^{'} }
\delta_{ {\bf{q}}, {\bf{q}}^{'} }\delta(t)
\]
\begin{equation}
+ (\frac{ v_{ {\bf{q}} } }{V})
\Lambda_{ {\bf{k}} }(-{\bf{q}})
\sum_{ {\bf{k}}^{''} }[\Lambda_{ {\bf{k}}^{''} }(-{\bf{q}})
\frac{ -i\langle T a^{t}_{ {\bf{k}}^{''} }({\bf{q}})
a^{\dagger}_{ {\bf{k}}^{'} }({\bf{q}}^{'}) \rangle }{\langle T 1 \rangle}
+ \Lambda_{ {\bf{k}}^{''} }({\bf{q}})
\frac{ -i\langle T a^{\dagger t}_{ {\bf{k}}^{''} }(-{\bf{q}})
a^{\dagger}_{ {\bf{k}}^{'} }({\bf{q}}^{'}) \rangle }{\langle T 1 \rangle}]
\end{equation}
\[
(i\frac{\partial}{\partial t} + \omega_{ {\bf{k}} }(-{\bf{q}}))
\frac{ -i\langle T a^{\dagger t}_{ {\bf{k}} }(-{\bf{q}})
a^{\dagger}_{ {\bf{k}}^{'} }({\bf{q}}^{'}) \rangle }{\langle T 1 \rangle}
\]
\begin{equation}
= -(\frac{ v_{ {\bf{q}} } }{V})
\Lambda_{ {\bf{k}} }({\bf{q}})
\sum_{ {\bf{k}}^{''} }[\Lambda_{ {\bf{k}}^{''} }({\bf{q}})
\frac{ -i\langle T a^{\dagger t}_{ {\bf{k}}^{''} }(-{\bf{q}})
a^{\dagger}_{ {\bf{k}}^{'} }({\bf{q}}^{'}) \rangle }{\langle T 1 \rangle}
+ \Lambda_{ {\bf{k}}^{''} }(-{\bf{q}})
\frac{ -i\langle T a^{t}_{ {\bf{k}}^{''} }({\bf{q}})
a^{\dagger}_{ {\bf{k}}^{'} }({\bf{q}}^{'}) \rangle }{\langle T 1 \rangle} ]
\end{equation}
The boundary conditions on these propagators may be written down as,
\begin{equation}
\frac{ -i\langle T a^{\dagger t}_{ {\bf{k}} }(-{\bf{q}})
a^{\dagger}_{ {\bf{k}}^{'} }({\bf{q}}^{'}) \rangle }{\langle T 1 \rangle}
 = \frac{ -i\langle T a^{\dagger (t - i\beta)}_{ {\bf{k}} }(-{\bf{q}})
a^{\dagger}_{ {\bf{k}}^{'} }({\bf{q}}^{'}) \rangle }{\langle T 1 \rangle}
\end{equation}
\begin{equation}
\frac{ -i\langle T a^{t}_{ {\bf{k}} }({\bf{q}})
a^{\dagger}_{ {\bf{k}}^{'} }({\bf{q}}^{'}) \rangle }{\langle T 1 \rangle}
 = \frac{ -i\langle T a^{(t - i\beta)}_{ {\bf{k}} }({\bf{q}})
a^{\dagger}_{ {\bf{k}}^{'} }({\bf{q}}^{'}) \rangle }{\langle T 1 \rangle}
\end{equation}
\begin{equation}
\delta(t) = (\frac{1}{-i \mbox{ }\beta})\sum_{n} exp(\omega_{n}t)
\end{equation}
\begin{equation}
\theta(t) = (\frac{1}{-i \mbox{ }\beta})\sum_{n} 
\frac{ exp(\omega_{n}t) }{\omega_{n}}
\end{equation}

The boundary conditions imply that we may write,
\begin{equation}
\frac{ -i\langle T a^{t}_{ {\bf{k}} }({\bf{q}})
a^{\dagger}_{ {\bf{k}}^{'} }({\bf{q}}^{'}) \rangle }{\langle T 1 \rangle}
 = \sum_{n}\mbox{ }exp(\omega_{n}t)\mbox{ }
\frac{ -i\langle T a^{n}_{ {\bf{k}} }({\bf{q}})
a^{\dagger}_{ {\bf{k}}^{'} }({\bf{q}}^{'}) \rangle }{\langle T 1 \rangle}
\end{equation}
\begin{equation}
\frac{ -i\langle T a^{\dagger t}_{ {\bf{k}} }(-{\bf{q}})
a^{\dagger}_{ {\bf{k}}^{'} }({\bf{q}}^{'}) \rangle }{\langle T 1 \rangle}
 = \sum_{n}\mbox{ }exp(\omega_{n}t)\mbox{ }
\frac{ -i\langle T a^{\dagger n}_{ {\bf{k}} }(-{\bf{q}})
a^{\dagger}_{ {\bf{k}}^{'} }({\bf{q}}^{'}) \rangle }{\langle T 1 \rangle}
\end{equation}
and, $ \omega_{n} = (2\mbox{ }\pi \mbox{ }n)/\beta $.  Thus,
\[
(i\omega_{n} - \omega_{ {\bf{k}} }({\bf{q}}))
\frac{ -i\langle T a^{n}_{ {\bf{k}} }({\bf{q}})
a^{\dagger}_{ {\bf{k}}^{'} }({\bf{q}}^{'}) \rangle }{\langle T 1 \rangle}
 = \frac{ \delta_{ {\bf{k}}, {\bf{k}}^{'} }
\delta_{ {\bf{q}}, {\bf{q}}^{'} } }{-i \mbox{ }\beta}
\]
\begin{equation}
+ (\frac{ v_{ {\bf{q}} } }{V})
\Lambda_{ {\bf{k}} }(-{\bf{q}})
\sum_{ {\bf{k}}^{''} }[\Lambda_{ {\bf{k}}^{''} }(-{\bf{q}})
\frac{ -i\langle T a^{n}_{ {\bf{k}}^{''} }({\bf{q}})
a^{\dagger}_{ {\bf{k}}^{'} }({\bf{q}}^{'}) \rangle }{\langle T 1 \rangle}
+ \Lambda_{ {\bf{k}}^{''} }({\bf{q}})
\frac{ -i\langle T a^{\dagger n}_{ {\bf{k}}^{''} }(-{\bf{q}})
a^{\dagger}_{ {\bf{k}}^{'} }({\bf{q}}^{'}) \rangle }{\langle T 1 \rangle}]
\end{equation}
\[
(i\omega_{n} + \omega_{ {\bf{k}} }(-{\bf{q}}))
\frac{ -i\langle T a^{\dagger n}_{ {\bf{k}} }(-{\bf{q}})
a^{\dagger}_{ {\bf{k}}^{'} }({\bf{q}}^{'}) \rangle }{\langle T 1 \rangle}
\]
\begin{equation}
= -(\frac{ v_{ {\bf{q}} } }{V})
\Lambda_{ {\bf{k}} }({\bf{q}})
\sum_{ {\bf{k}}^{''} }[\Lambda_{ {\bf{k}}^{''} }({\bf{q}})
\frac{ -i\langle T a^{\dagger n}_{ {\bf{k}}^{''} }(-{\bf{q}})
a^{\dagger}_{ {\bf{k}}^{'} }({\bf{q}}^{'}) \rangle }{\langle T 1 \rangle}
+ \Lambda_{ {\bf{k}}^{''} }(-{\bf{q}})
\frac{ -i\langle T a^{n}_{ {\bf{k}}^{''} }({\bf{q}})
a^{\dagger}_{ {\bf{k}}^{'} }({\bf{q}}^{'}) \rangle }{\langle T 1 \rangle} ]
\end{equation}
Define,
\begin{equation}
\sum_{ {\bf{k}} } \Lambda_{ {\bf{k}} } (-{\bf{q}})
\frac{ -i\langle T a^{n}_{ {\bf{k}} }({\bf{q}})
a^{\dagger}_{ {\bf{k}}^{'} }({\bf{q}}^{'}) \rangle }{\langle T 1 \rangle} 
 = G_{1}( {\bf{q}}, {\bf{k}}^{'}, {\bf{q}}^{'}; n)
\end{equation}
\begin{equation}
\sum_{ {\bf{k}} } \Lambda_{ {\bf{k}} } ({\bf{q}})
\frac{ -i\langle T a^{\dagger n}_{ {\bf{k}} }(-{\bf{q}})
a^{\dagger}_{ {\bf{k}}^{'} }({\bf{q}}^{'}) \rangle }{\langle T 1 \rangle} 
 = G_{2}( {\bf{q}}, {\bf{k}}^{'}, {\bf{q}}^{'}; n)
\end{equation}
Multiplying the above equations with $ \Lambda_{ {\bf{k}} }(-{\bf{q}}) $ and
summing over $ {\bf{k}} $ one arrives at simple formulas for
 $ G_{1} $ and $ G_{2} $.
\[
G_{1}( {\bf{q}}, {\bf{k}}^{'}, {\bf{q}}^{'}; n) = 
\Lambda_{ {\bf{k}}^{'} }(-{\bf{q}})
\frac{ \delta_{ {\bf{q}}, {\bf{q}}^{'} } }
{ -i \mbox{ }\beta( i\omega_{n} - \omega_{ {\bf{k}}^{'} }({\bf{q}}) ) }
\]
\begin{equation}
+ f_{n}({\bf{q}})[G_{1}( {\bf{q}}, {\bf{k}}^{'}, {\bf{q}}^{'}; n)
+ G_{2}( {\bf{q}}, {\bf{k}}^{'}, {\bf{q}}^{'}; n) ]
\end{equation}
and,
\begin{equation}
G_{2}( {\bf{q}}, {\bf{k}}^{'}, {\bf{q}}^{'}; n) =
f^{*}_{n}(-{\bf{q}})[ G_{1}( {\bf{q}}, {\bf{k}}^{'}, {\bf{q}}^{'}; n)
+ G_{2}( {\bf{q}}, {\bf{k}}^{'}, {\bf{q}}^{'}; n) ]
\end{equation}
and,
\begin{equation}
f_{n}({\bf{q}}) = (\frac{ v_{ {\bf{q}} } }{V})
\sum_{ {\bf{k}} }\frac{ \Lambda_{ {\bf{k}} }(-{\bf{q}}) }
{ ( i\omega_{n} - \omega_{ {\bf{k}} }({\bf{q}}) ) }
\end{equation}
In 1D and 3D an analytical solution for $ f_{n}({\bf{q}}) $ is possible.
In 1D we have,
\begin{equation}
f_{n}(q) = v_{q}(-\frac{m}{q})(\frac{1}{2\pi})
ln ( \frac{ k_{f} - \frac{ m \mbox{ }i\omega_{n} }{q} + \frac{q}{2} }
{ -k_{f} - \frac{ m \mbox{ }i\omega_{n} }{q} + \frac{q}{2}  } )
 - \theta(b-a)\mbox{ }v_{q}\mbox{ }(\frac{1}{2\pi})(-\frac{m}{q})
ln( \frac{ b - \frac{ m \mbox{ }i\omega_{n} }{q} + \frac{q}{2} }
{ a - \frac{ m \mbox{ }i\omega_{n} }{q} + \frac{q}{2}  } )
\end{equation}
where, $ b = min(k_{f}, k_{f} - q) $ and $ a = max(-k_{f}, -k_{f}-q) $. 
The logarithm is to be interpreted as the principal value,  that
is, $ Im(ln(z)) = Arg(z) = tan^{-1}(y/x) $, $ x = Re(z), y = Im(z) $, 
 and $ Re(ln(z)) = ln(r) $ and, $ r = (x^{2} + y^{2})^{\frac{1}{2}} $.
 The nice thing about this definition (not shared by
  non-principal value defns.) is that $ (ln(z))^{*} = ln(z^{*}) $.
 This is because $ -\pi \leq Arg(z) = tan^{-1}(y/x) \leq \pi $.
This may easily be solved and the final formulas for the Bose propagators
 may be written down as,
\[
G_{ 2 }({\bf{q}}, {\bf{k}}^{'}, {\bf{q}}^{'}; n) =
  \frac{ f^{*}_{n}(-{\bf{q}}) }{(1- f^{*}_{n}(-{\bf{q}}))}
G_{1} ({\bf{q}}, {\bf{k}}^{'}, {\bf{q}}^{'}; n)
\]
\[
 G_{1}({\bf{q}}, {\bf{k}}^{'}, {\bf{q}}^{'}; n) +
 G_{2}({\bf{q}}, {\bf{k}}^{'}, {\bf{q}}^{'}; n)
 = G_{1}({\bf{q}}, {\bf{k}}^{'}, {\bf{q}}^{'}; n) /( 1 - f^{*}_{n}(-{\bf{q}}) ) 
\]
\begin{equation}
G_{1}({\bf{q}}, {\bf{k}}^{'}, {\bf{q}}^{'}; n)
 = (\frac{1}{-i \mbox{ }\beta})
\frac{ (1-f_{n}^{*}(-{\bf{q}}) ) \Lambda_{ {\bf{k}}^{'} }(-{\bf{q}})
\delta_{ {\bf{q}}, {\bf{q}}^{'} } }
{(1- f_{n}^{*}(-{\bf{q}}) - f_{n}({\bf{q}}) )
 ( i\omega_{n} - \omega_{ {\bf{k}}^{'} }({\bf{q}}) ) }
\end{equation}
\begin{equation}
G_{2}({\bf{q}}, {\bf{k}}^{'}, {\bf{q}}^{'}; n)
 = (\frac{1}{-i \mbox{ }\beta})
\frac{ f_{n}^{*}(-{\bf{q}}) \Lambda_{ {\bf{k}}^{'} }(-{\bf{q}})
\delta_{ {\bf{q}}, {\bf{q}}^{'} } }
{(1- f_{n}^{*}(-{\bf{q}}) - f_{n}({\bf{q}}) )
 ( i\omega_{n} - \omega_{ {\bf{k}}^{'} }({\bf{q}}) ) }
\end{equation}
\[
G_{1}({\bf{q}}, {\bf{k}}^{'}, {\bf{q}}^{'}; n)
 + G_{2}({\bf{q}}, {\bf{k}}^{'}, {\bf{q}}^{'}; n)
 = (\frac{1}{-i \mbox{ }\beta})\frac{ \Lambda_{ {\bf{k}}^{'} }(-{\bf{q}})
\delta_{ {\bf{q}}, {\bf{q}}^{'} } }
{(1- f_{n}^{*}(-{\bf{q}}) - f_{n}({\bf{q}}) )
 ( i\omega_{n} - \omega_{ {\bf{k}}^{'} }({\bf{q}}) ) }
\]

\[
\frac{ -i\langle T a^{n}_{ {\bf{k}} }({\bf{q}})
a^{\dagger}_{ {\bf{k}}^{'} }({\bf{q}}^{'}) \rangle }{ \langle T 1 \rangle }
 = \frac{ \delta_{ {\bf{k}}, {\bf{k}}^{'} } \delta_{ {\bf{q}}, {\bf{q}}^{'} } }
{-i\mbox{ }\beta( i\omega_{n} - \omega_{ {\bf{k}} }({\bf{q}}) )}
\]
\begin{equation}
+ (\frac{1}{-i \mbox{ }\beta})
( \frac{ v_{ {\bf{q}} } }{V} )\frac{ \Lambda_{ {\bf{k}} }(-{\bf{q}}) }
{(i\omega_{ n } - \omega_{ {\bf{k}} }({\bf{q}}) ) }
\frac{ \Lambda_{ {\bf{k}}^{'} }(-{\bf{q}})
\delta_{ {\bf{q}}, {\bf{q}}^{'} } }
{(1- f_{n}^{*}(-{\bf{q}}) - f_{n}({\bf{q}}) )
 ( i\omega_{n} - \omega_{ {\bf{k}}^{'} }({\bf{q}}) ) }
\end{equation}
also,
\begin{equation}
\frac{ -i\langle T a^{\dagger n}_{ {\bf{k}} }(-{\bf{q}})
a^{\dagger}_{ {\bf{k}}^{'} }({\bf{q}}^{'}) \rangle }{ \langle T 1 \rangle }
 = 
-(\frac{1}{-i \mbox{ } \beta})
( \frac{ v_{ {\bf{q}} } }{V} )
\frac{ \Lambda_{ {\bf{k}} }({\bf{q}}) }
{(i\omega_{ n } + \omega_{ {\bf{k}} }(-{\bf{q}}) ) }
\frac{ \Lambda_{ {\bf{k}}^{'} }(-{\bf{q}})
\delta_{ {\bf{q}}, {\bf{q}}^{'} } }
{(1- f_{n}^{*}(-{\bf{q}}) - f_{n}({\bf{q}}) )
 ( i\omega_{n} - \omega_{ {\bf{k}}^{'} }({\bf{q}}) ) }
\end{equation}
The zero temperature correlation function of significance here is,
\begin{equation}
-i\langle a^{\dagger}_{ {\bf{k}}^{'} }({\bf{q}}^{'})
a_{ {\bf{k}} }({\bf{q}}) \rangle
\end{equation}
This may be obtained from the above formulas as,
\begin{equation}
-i\langle a^{\dagger}_{ {\bf{k}}^{'} }({\bf{q}}^{'})
a_{ {\bf{k}} }({\bf{q}}) \rangle
 = -(\frac{v_{ {\bf{q}} } }{V})
\Lambda_{ {\bf{k}} }(-{\bf{q}})\Lambda_{ {\bf{k}}^{'} }(-{\bf{q}})
\delta_{ {\bf{q}}, {\bf{q}}^{'} }
\int_{C} \frac{d\omega}{2\pi \mbox{ }i} 
\frac{1}{( i\omega - \omega_{ {\bf{k}} }({\bf{q}}) )
( i\omega - \omega_{ {\bf{k}}^{'} }({\bf{q}}) )
 ( 1 - f_{n}^{*}(-{\bf{q}}) - f_{n}({\bf{q}}) ) }
\end{equation}
where $ C $ is the positively oriented contour
 that encloses the upper half plane. 
Thus the problem now boils down to computing all the zeros
 of $ ( 1 - f_{n}^{*}(-{\bf{q}}) - f_{n}({\bf{q}}) )  $
 that have positive imaginary parts. In 1D, we may 
 proceed as follows,
\[
1 - f^{*}_{n}(-q) - f_{n}(q)
 = 1 - v_{q}(-\frac{m}{q})(\frac{1}{2\pi})
ln[\frac{k_{f} - \frac{m \mbox{ }i\omega}{q} + \frac{q}{2} }
{ -k_{f} - \frac{m \mbox{ }i\omega}{q} + \frac{q}{2} }]
\]
\[
+ \theta(b-a) v_{q}(\frac{1}{2\pi})
(-\frac{m}{q})ln[\frac{b - \frac{m \mbox{ }i\omega}{q} + \frac{q}{2} }
{ a - \frac{m \mbox{ }i\omega}{q} + \frac{q}{2} }]
\]
\[
 - v_{q}(\frac{m}{q})(\frac{1}{2\pi})
ln[\frac{k_{f} - \frac{m \mbox{ }i\omega}{q} - \frac{q}{2} }
{ -k_{f} - \frac{m \mbox{ }i\omega}{q} - \frac{q}{2} }]
\]
\begin{equation}
+ \theta(b^{'}-a^{'}) v_{q}(\frac{1}{2\pi})
(\frac{m}{q})ln[\frac{b^{'} - \frac{m \mbox{ }i\omega}{q} - \frac{q}{2} }
{ a^{'} - \frac{m \mbox{ }i\omega}{q} - \frac{q}{2} }]
\end{equation}
where $ a = max(- k_{f}, - k_{f} - q) $, $ b = min(k_{f}, k_{f} - q) $
 and $ a^{'} = max(- k_{f}, - k_{f} + q) $, 
$ b^{'} = min(k_{f}, k_{f} + q) $. There are several regions of interest.
\newline
(A) $  0 \leq q \leq k_{f}  $
 This means, $  0 \geq -q \geq -k_{f}  $
 or $ k_{f} \geq k_{f}-q \geq 0  $, $  -k_{f}  \geq -k_{f}-q \geq -2k_{f}  $,
 $ k_{f} \leq k_{f} + q \leq 2k_{f} $, $  -k_{f} \leq -k_{f} + q \leq 0  $
 Thus $ a = -k_{f} $, $ b = k_{f} - q $, $ a^{'} = -k_{f} + q $, 
 $ b^{'} = k_{f} $. $ \theta(b-a) = 1 $, $ \theta(b^{'}-a^{'}) = 1 $
Therefore,
\[
1 - f^{*}_{n}(-q) - f_{n}(q)
 = 1 + v_{q}(\frac{m}{q})(\frac{1}{2\pi})
ln[\frac{k_{f} - \frac{m \mbox{ }i\omega}{q} + \frac{q}{2} }
{ k_{f} - \frac{m \mbox{ }i\omega}{q} - \frac{q}{2} }]
\]
\begin{equation}
+ v_{q}(\frac{1}{2\pi})
(\frac{m}{q})ln[\frac{ -k_{f} - \frac{m \mbox{ }i\omega}{q} - \frac{q}{2} }
{ - k_{f}  - \frac{m \mbox{ }i\omega}{q} + \frac{q}{2} }]
\end{equation}
\begin{equation}
1 - f^{*}_{n}(-q) - f_{n}(q) = 
1 + v_{q}(\frac{1}{2\pi})
(\frac{m}{q})ln[ \frac{ (k_{f}+q/2)^{2} + (\frac{m \mbox{ }\omega}{q})^{2} }
{ (k_{f} - q/2)^{2} + (\frac{m \mbox{ }\omega}{q})^{2} } ]
 = 0
\end{equation}
\begin{equation}
ln[ \frac{ (k_{f}+q/2)^{2} + (\frac{m \mbox{ }\omega}{q})^{2} }
{ (k_{f} - q/2)^{2} + (\frac{m \mbox{ }\omega}{q})^{2} } ]
 = -(\frac{ 2\mbox{ }\pi\mbox{ }q }{m})(\frac{1}{v_{q}})
\end{equation}
\begin{equation}
\frac{ (k_{f}+q/2)^{2} + (\frac{m \mbox{ }\omega}{q})^{2} }
{ (k_{f} - q/2)^{2} + (\frac{m \mbox{ }\omega}{q})^{2} }
  = exp(-(\frac{ 2\mbox{ }\pi\mbox{ }q }{m})(\frac{1}{v_{q}}))
\end{equation}
\begin{equation}
(\frac{m \mbox{ }\omega}{q})^{2}
 = -[(k_{f}+q/2)^{2} - (k_{f} - q/2)^{2} 
exp(-(\frac{ 2\mbox{ }\pi\mbox{ }q }{m})(\frac{1}{v_{q}})) ]
/[1- exp(-(\frac{ 2\mbox{ }\pi\mbox{ }q }{m})(\frac{1}{v_{q}}))]
\end{equation}
 We want to find a root of this that has a positive
 imaginary part.
\begin{equation}
\omega = i\mbox{ }(\frac{q}{m})
\sqrt{ \frac{ (k_{f}+q/2)^{2} - (k_{f} - q/2)^{2}
exp(-(\frac{ 2\mbox{ }\pi\mbox{ }q }{m})(\frac{1}{v_{q}})) }
{1- exp(-(\frac{ 2\mbox{ }\pi\mbox{ }q }{m})(\frac{1}{v_{q}}))} }
\end{equation}
 The next case is,
\newline
 (B) $ -k_{f} \leq q \leq 0 $.
 In this case, $ 0 \leq k_{f} + q \leq k_{f} $,
 $ -2\mbox{ }k_{f} \leq -k_{f} + q \leq -k_{f} $,
 $ k_{f} \geq -q \geq 0 $, $ 2\mbox{ }k_{f} \geq k_{f}-q \geq k_{f} $,
 $ 0 \geq -k_{f}-q \geq -k_{f} $, $ a = -k_{f} - q $,
 $ b = k_{f} $, $ a^{'} = -k_{f} $, $ b^{'} = k_{f} + q $.
 Thus, $ \theta(b-a) = 1 $, $ \theta(b^{'} - a^{'}) = 1 $
 For case (B)
\[
1 - f^{*}_{n}(-q) - f_{n}(q)
 = 1 + v_{q}(\frac{m}{q})(\frac{1}{2\pi})
ln[\frac{-k_{f} - \frac{m \mbox{ }i\omega}{q} - \frac{q}{2} }
{ -k_{f} - \frac{m \mbox{ }i\omega}{q} + \frac{q}{2} }]
\]
\begin{equation}
+ v_{q}(\frac{1}{2\pi})
(\frac{m}{q})ln[\frac{k_{f} - \frac{m \mbox{ }i\omega}{q} + \frac{q}{2} }
{  k_{f} - \frac{m \mbox{ }i\omega}{q} - \frac{q}{2} }]
\end{equation}
Thus the pole for both cases (A) and (B) is given by,
\begin{equation}
\omega = i\mbox{ }(\frac{|q|}{m})
\sqrt{ \frac{ (k_{f}+q/2)^{2} - (k_{f} - q/2)^{2}
exp(-(\frac{ 2\mbox{ }\pi\mbox{ }q }{m})(\frac{1}{v_{q}})) }
{ 1- exp(-(\frac{ 2\mbox{ }\pi\mbox{ }q }{m})(\frac{1}{v_{q}})) } }
\label{ROOT}
\end{equation}
The quantity under the square root is positive in both cases.
Third, we have the case,
\newline
(C) $ k_{f} \leq q \leq 2\mbox{ }k_{f} $ , this means,
 $ -k_{f} \geq -q \geq -2\mbox{ }k_{f} $, also,
 $ 2\mbox{ }k_{f} \leq k_{f} + q \leq 3\mbox{ }k_{f} $,
 $ 0 \leq -k_{f} + q \leq k_{f} $, $ 0 \geq k_{f}-q \geq -k_{f} $,
 $ -2\mbox{ }k_{f} \geq -k_{f}-q \geq -3\mbox{ }k_{f} $.
 $ a = -k_{f} $, $ b = k_{f} - q $, $ a^{'} = -k_{f} + q $,
 $ b^{'} = k_{f} $. 
 $ \theta(b-a) = 1 $ and $ \theta(b^{'}-a^{'}) = 1 $
 This leads to the same root Eq.(~\ref{ROOT}). In case
\newline
(D) $ -2\mbox{ }k_{f} \leq q \leq -k_{f} $, 
$ -k_{f} \leq k_{f}+ q \leq 0 $, 
$ -3\mbox{ }k_{f} \leq -k_{f}+q \leq -2\mbox{ }k_{f} $,
$ 2\mbox{ }k_{f} \geq -q \geq k_{f} $, 
 $ 3\mbox{ }k_{f} \geq k_{f}-q \geq 2\mbox{ }k_{f} $,
 $ k_{f} \geq -k_{f}-q \geq 0 $. $ a = -k_{f}-q $,
 $ b = k_{f} $, $ a^{'} = -k_{f} $, $ b^{'} = k_{f}+ q $.
 $ \theta(b-a) = 1 $ and $ \theta(b^{'}-a^{'}) = 1 $.
 In this case also we find the root given by Eq.(~\ref{ROOT}).
(E)  $ 2\mbox{ }k_{f} \leq q < \infty $, 
 $ -2\mbox{ }k_{f} \geq -q > -\infty $, 
 $ 3\mbox{ }k_{f} \leq k_{f}+q < \infty $,
 $ k_{f} \leq -k_{f}+q < \infty $,
 $ -k_{f} \geq k_{f}-q > -\infty $,
 $ -3\mbox{ }k_{f} \geq -k_{f}-q > -\infty $.
 $ a = -k_{f} $, $ b =  k_{f}-q $, $ a^{'} = -k_{f}+q $
 $ b^{'} = k_{f} $. $ \theta(b-a) = 0 $, 
 $ \theta(b^{'}-a^{'}) = 0 $.  This case is somewhat different.
\[
1 - f^{*}_{n}(-q) - f_{n}(q)
 = 1 + v_{q}(\frac{m}{q})(\frac{1}{2\pi})
ln[\frac{k_{f} - \frac{m \mbox{ }i\omega}{q} + \frac{q}{2} }
{ -k_{f} - \frac{m \mbox{ }i\omega}{q} + \frac{q}{2} }]
\]
\begin{equation}
 - v_{q}(\frac{m}{q})(\frac{1}{2\pi})
ln[\frac{k_{f} - \frac{m \mbox{ }i\omega}{q} - \frac{q}{2} }
{ -k_{f} - \frac{m \mbox{ }i\omega}{q} - \frac{q}{2} }]
\end{equation}
This when solved gives the same result as Eq.(~\ref{ROOT}).
 So too does the final case (F) which is
\newline
(F) $ -\infty < q \leq -2 \mbox{ }k_{f} $,
 $ -\infty < k_{f}+q \leq -k_{f} $,
 $ -\infty < -k_{f}+q \leq -3 \mbox{ }k_{f} $,
 $ \infty > -q \geq 2 \mbox{ }k_{f} $,
 $ \infty > k_{f}-q \geq 3 \mbox{ }k_{f} $,
 $ \infty > -k_{f}-q \geq k_{f} $. 
This case also leads to the same result namely, Eq.(~\ref{ROOT}).  
Therefore the final result may be written as,
\begin{equation}
\langle a^{\dagger}_{ {\bf{k}}^{'} }({\bf{q}}^{'})a_{ {\bf{k}} }({\bf{q}})
 \rangle
 = (\frac{1}{V})\frac
{ \Lambda_{k}(-q)\Lambda_{k^{'}}(-q) \delta_{ q, q^{'} } }
{ (\omega_{R}(q)+ \omega_{k}(q))(\omega_{R}(q)+ \omega_{k^{'}}(q))
(\frac{m}{q^{2}})(\frac{1}{2 \pi k_{f}})2(\frac{m}{q})^{2}\omega_{R}(q)
(cosh(\lambda(q))-1) }
\end{equation}
Here we may write,
\begin{equation}
\lambda(q) = (\frac{2 \pi q}{m})(\frac{1}{v_{q}})
\end{equation}
\begin{equation}
\omega_{R}(q) = (\frac{ |q| }{m})
\sqrt{ \frac{ (k_{f} + q/2)^{2}  - (k_{f} - q/2)^{2}exp(-\lambda(q)) }
{ 1 - exp(-\lambda(q)) } }
\end{equation}

\newpage
In other words,
\[
\langle c^{\dagger}_{ k }c_{ k } \rangle
 = n_{F}(k) + 
(2\pi k_{f})
\int_{-\infty}^{+\infty} \mbox{ }\frac{ dq_{1} }{2\pi}\mbox{ }
\frac{ \Lambda_{ k - q_{1}/2 }(-q_{1}) }
{ 2\omega_{R}(q_{1})(\omega_{R}(q_{1}) + \omega_{k - q_{1}/2}(q_{1}))^{2}
(\frac{ m^{3} }{q_{1}^{4}})( cosh(\lambda(q_{1})) - 1 ) }
\]
\begin{equation}
- (2\pi k_{f})
\int_{-\infty}^{+\infty} \mbox{ }\frac{ dq_{1} }{2\pi}\mbox{ }
\frac{ \Lambda_{ k + q_{1}/2 }(-q_{1}) }
{ 2\omega_{R}(q_{1})(\omega_{R}(q_{1}) + \omega_{k + q_{1}/2}(q_{1}))^{2}
(\frac{ m^{3} }{q_{1}^{4}})( cosh(\lambda(q_{1})) - 1 ) }
\end{equation}
 In order to see how good the present theory is, it is desirable to
 compare these results with the Calogero-Sutherland model or
 more specifically with the spin-spin correlation function of the
 Haldane-Shastry model. This is given by \cite{Lesage},
 $ \rho = 1/2 $, $ \alpha = 2 $, $ m = 1 $, $ k_{f} = \pi/2 $.
\[
\langle 0 | \Psi^{\dagger}(x,0)\Psi(x^{'},0) | 0 \rangle
 = \int_{-\infty}^{+\infty} \mbox{ }\frac{ dk }{2\pi}\mbox{ }
 exp(i \mbox{ }k\mbox{ }(x^{'}-x))
\langle c^{\dagger}_{k}c_{k} \rangle 
\]
\[
 = (\frac{1}{4})\frac{ (\Gamma(3/2))^{2} }
{ (\Gamma(1/2))^{2} (\Gamma(1))^{2} }
\int_{-1}^{+1} \mbox{ } dv_{1} \mbox{ }
\int_{-1}^{+1} \mbox{ } dv_{2} \mbox{ }
(1-v^{2}_{1})^{ -\frac{1}{2} }
(1-v^{2}_{2})^{ -\frac{1}{2} } |v_{1} - v_{2}|
\]
\begin{equation}
exp(i\mbox{ }\frac{\pi}{2}(x-x^{'})\mbox{ }v_{1})
exp(i\mbox{ }\frac{\pi}{2}(x-x^{'})\mbox{ }v_{2})
\end{equation}
\[
 = 4\mbox{ }\frac{ (\Gamma(3/2))^{2} }
{ (\Gamma(1/2))^{2} (\Gamma(1))^{2} }
\sum_{n = 0}(-)^{n}\frac{ \pi^{2n}(x-x^{'})^{2n} }{ (2\mbox{ }n)! }
 \frac{1}{(2 \mbox{ }n + 1)^{2}}
\]
For this we have to use the interaction given by 
($ V(x) = \frac{\alpha(\alpha-1)}{x^{2}} $),
\begin{equation}
v_{q} = -\alpha(\alpha-1) \mbox{ }(\pi |q|)
\end{equation}

\subsection{ Qualitative Conclusions }
 It is clear that one of the features of
 the Luttinger liquid that is a result of the Mattis-Leib solution
 is absent, namely the discontinuous dependence of the momentum
 distribution on the coupling strength as the latter goes to zero.
 This may therefore
 be an artifact of the Luttinger model and not a generic feature
 of all 1D systems. Also the highly nonperturbative manner
 in which the momentum distribution of the interacting system
 approaches the noninteracting one as the coupling goes to
 zero may be seen quite easily. The other point that may be seen
 is that $ \langle \rho_{ {q} \neq 0  } \rangle = 0 $ indicating that
 there is no Wigner crystallization at any density.
 Furthermore, the vanishing of the quasiparticle residue is seen
 only for sufficiently large values of the repulsion between the
 fermions. Thus the conventional view that the fermi surface is destroyed
 for arbitrarily weak repulsion in case of 1d fermions does not seem
 to hold up in this case. 

\subsection{Momentum Distribution in 3D }
For this one has to compute the expressions in 3D:
\[
f_{n}({\bf{q}})
 = \frac{ v_{ {\bf{q}} } }{(2\pi)^{2}}
\int_{0}^{k_{f}} \mbox{ }dk \mbox{ }k^{2}
\mbox{ }
\{
(-\frac{m}{k|q|})
ln(\frac{ i\omega_{n} - \epsilon_{ {\bf{q}} } - \frac{ k |q| }{m} }
{ i\omega_{n} - \epsilon_{ {\bf{q}} } + \frac{ k |q| }{m} })
\]
\[
 + (\frac{m}{k|q|})\theta(k_{F} - |q| - k )
ln(\frac{ i\omega_{n} - \epsilon_{ {\bf{q}} } - \frac{ k |q| }{m} }
{ i\omega_{n} - \epsilon_{ {\bf{q}} } + \frac{ k |q| }{m} })
\]
\begin{equation}
+ (\frac{m}{k|q|})\theta(|q| + k - k_{f})
\theta(k_{f}^{2} - (k - |q|)^{2})
ln(\frac{ i\omega_{n} - \epsilon_{ k_{f} } + \epsilon_{ {\bf{k}} } }
{ i\omega_{n} - \epsilon_{ {\bf{q}} } + \frac{ k |q| }{m} })
\}
\end{equation}
\newpage
This may be split into several cases
\newline
(A) $ 0 < |{\bf{q}}| < k_{f} $
\[
f_{n}({\bf{q}})
 = \frac{v_{ {\bf{q}} }}{(2\pi)^{2}}
(\frac{m}{|{\bf{q}}|})
\{
-\int_{0}^{k_{f}}dk \mbox{ }k\mbox{ }
ln(\frac{i\omega_{n} - \epsilon_{ {\bf{q}} } - \frac{ k |{\bf{q}}| }{m} }
{ i\omega_{n} - \epsilon_{ {\bf{q}} } + \frac{ k |{\bf{q}}| }{m} })
\]
\[
+ \int_{0}^{k_{f}-|q|}dk \mbox{ }k\mbox{ }
ln(\frac{i\omega_{n} - \epsilon_{ {\bf{q}} } - \frac{ k |{\bf{q}}| }{m} }
{ i\omega_{n} - \epsilon_{ {\bf{q}} } + \frac{ k |{\bf{q}}| }{m} })
\]
\begin{equation}
+ \int_{k_{F}-|{\bf{q}}|}^{k_{f}}dk \mbox{ }k\mbox{ }
ln(\frac{i\omega_{n} - \epsilon_{ k_{F} } + \epsilon_{ {\bf{k}} } }
{ i\omega_{n} - \epsilon_{ {\bf{q}} } + \frac{ k |{\bf{q}}| }{m} }) \}
\end{equation}
\newline
(B) $ k_{f} < |q| < 2\mbox{ }k_{f} $
\[
f_{n}({\bf{q}})
 = \frac{ v_{ {\bf{q}} } }{(2\pi)^{2}}
(\frac{m}{|{\bf{q}}|})
\{
-\int_{0}^{k_{f}}
dk \mbox{ }k \mbox{ }
ln(\frac{ i\omega_{n} - \epsilon_{ {\bf{q}} } - \frac{ k|q|}{m} }
{ i\omega_{n} - \epsilon_{ {\bf{q}} } + \frac{ k|q|}{m} })
\]
\begin{equation}
+ \int_{|q|-k_{f}}^{k_{f}}
dk \mbox{ }k \mbox{ }
ln(\frac{ i\omega_{n} - \epsilon_{ k_{f} } + \epsilon_{ {\bf{k}} } }
{ i\omega_{n} - \epsilon_{ {\bf{q}} } + \frac{ k|q|}{m} }) \}
\end{equation}
\newline
(C) $ |q| > 2\mbox{ }k_{f} $
\newline
\begin{equation}
f_{n}({\bf{q}})
 = -\frac{ v_{ {\bf{q}} } }{(2\pi)^{2}}
(\frac{m}{|{\bf{q}}|})
\int_{0}^{k_{f}}
dk \mbox{ }k \mbox{ }
ln(\frac{ i\omega_{n} - \epsilon_{ {\bf{q}} } - \frac{ k|q|}{m} }
{ i\omega_{n} - \epsilon_{ {\bf{q}} } + \frac{ k|q|}{m} })
\end{equation}
\newline
(A)
\newline
\[
f_{n}({\bf{q}})
 = \frac{ v_{ {\bf{q}} } }{(2\pi)^{2}}
(\frac{m}{|q|})
[-\frac{1}{2}k_{f}^{2}
ln(i\omega_{n} - \epsilon_{ {\bf{q}} } - \frac{k_{f}|q|}{m})
+ \frac{1}{2}(k_{f}-|q|)^{2}
ln(i\omega_{n} + \epsilon_{ {\bf{q}} } - \frac{k_{f}|q|}{m})
\]
\[
+ (\frac{m^{2}}{2\mbox{ }q^{2}})
\{ \frac{1}{2}(i\omega_{n} - \epsilon_{ {\bf{q}} }
 - \frac{ k_{f} |q| }{m})^{2}
 - \frac{1}{2}(i\omega_{n} + \epsilon_{ {\bf{q}} }
 - \frac{ k_{f} |q| }{m})^{2}
\]
\[
+ (i\omega_{n} - \epsilon_{ {\bf{q}} })^{2}
ln(\frac{ i\omega_{n} - \epsilon_{ {\bf{q}} }
 - \frac{ k_{f} |q| }{m} }
{ i\omega_{n} + \epsilon_{ {\bf{q}} }
 - \frac{ k_{f} |q| }{m} })
\]
\[
+ (\frac{2|q|^{2}}{m})
(i\omega_{n}-\epsilon_{ {\bf{q}} })
\}
\]
\begin{equation}
+ m\mbox{ }(i\omega_{n})ln(i\omega_{n}) - m\mbox{ }i\omega_{n}
 - m(\epsilon_{ {\bf{q}} } - \frac{k_{f} |q|}{m} + i\omega_{n})
ln(\epsilon_{ {\bf{q}} } - \frac{k_{f} |q|}{m} + i\omega_{n})
 + m(\epsilon_{ {\bf{q}} } - \frac{k_{f} |q|}{m} + i\omega_{n}) ]
\end{equation}
\newline
(B)
\newline
\[
f_{n}({\bf{q}})
 = \frac{ v_{ {\bf{q}} } }{(2\pi)^{2}}
(\frac{m}{|q|})
[-\frac{1}{2}k_{f}^{2}
ln(i\omega_{n} - \epsilon_{ {\bf{q}} } - \frac{k_{f}|q|}{m})
+ \frac{1}{2}(k_{f}-|q|)^{2}
ln(i\omega_{n} + \epsilon_{ {\bf{q}} } - \frac{k_{f}|q|}{m})
\]
\[
+ (\frac{m^{2}}{2\mbox{ }q^{2}})
\{ \frac{1}{2}(i\omega_{n} - \epsilon_{ {\bf{q}} }
 - \frac{ k_{f} |q| }{m})^{2}
 - \frac{1}{2}(i\omega_{n} + \epsilon_{ {\bf{q}} }
 - \frac{ k_{f} |q| }{m})^{2}
\]
\[
+ (i\omega_{n} - \epsilon_{ {\bf{q}} })^{2}
ln(\frac{ i\omega_{n} - \epsilon_{ {\bf{q}} }
 - \frac{ k_{f} |q| }{m} }
{ i\omega_{n} + \epsilon_{ {\bf{q}} }
 - \frac{ k_{f} |q| }{m} })
\]
\[
+ (\frac{2|q|^{2}}{m})
(i\omega_{n}-\epsilon_{ {\bf{q}} })
\}
\]
\begin{equation}
+ m\mbox{ }(i\omega_{n})ln(i\omega_{n}) - m\mbox{ }i\omega_{n}
 - m(\epsilon_{ {\bf{q}} } - \frac{k_{f} |q|}{m} + i\omega_{n})
ln(\epsilon_{ {\bf{q}} } - \frac{k_{f} |q|}{m} + i\omega_{n})
 + m(\epsilon_{ {\bf{q}} } - \frac{k_{f} |q|}{m} + i\omega_{n}) ]
\end{equation}
\newline
(C)
\newline
\[
f_{n}({\bf{q}}) = \frac{ v_{ {\bf{q}} } }{(2\pi)^{2}}
(\frac{m}{|q|})
[ \frac{1}{2}k_{f}^{2}
ln(\frac{i\omega_{n} - \epsilon_{ {\bf{q}} } + \frac{ k_{f} |q| }{m} }
{ i\omega_{n} - \epsilon_{ {\bf{q}} } - \frac{ k_{f} |q| }{m} })
\]
\[
-\frac{m^{2}}{2|q|^{2}}
\{
\frac{1}{2}(i\omega_{n} - \epsilon_{ {\bf{q}} } + \frac{ k_{f} |q|}{m})^{2}
 - \frac{1}{2}(i\omega_{n} - \epsilon_{ {\bf{q}} } - \frac{ k_{f} |q|}{m})^{2}
\]
\begin{equation}
+ (i\omega_{n} - \epsilon_{ {\bf{q}} })^{2}
ln(\frac{ i\omega_{n} - \epsilon_{ {\bf{q}} } + \frac{ k_{f} |q|}{m} }
{ i\omega_{n} - \epsilon_{ {\bf{q}} } - \frac{ k_{f} |q|}{m} } )
 - \frac{4\mbox{ }k_{f}|q|}{m}(i\omega_{n} - \epsilon_{ {\bf{q}} })
\} ]
\end{equation}

\section{Fermions at Finite Temperature}

 It should be clear from the previous sections that fermions at finite
 temp. is a major headache. A straightforward generalisation is not
 working out for unknown reasons. In this section we shall not distinguish
 between $ N $ and $ \langle N \rangle $, things are complicated
 enough as it is !
 Take for example, the expectation
 value of the number density at finite temerature,
\[
\langle c^{\dagger}_{ {\bf{k}} }c_{ {\bf{k}} } \rangle
 = n_{F}({\bf{k}})
 + \sum_{ {\bf{q}}_{1} }
 \Lambda_{ {\bf{k}} - {\bf{q}}_{1}/2 }(-{\bf{q}}_{1})
 \langle a^{\dagger}_{ {\bf{k}} - {\bf{q}}_{1}/2 }({\bf{q}}_{1})
 a_{ {\bf{k}} - {\bf{q}}_{1}/2 }({\bf{q}}_{1}) \rangle
\]
\begin{equation}
 - \sum_{ {\bf{q}}_{1} }
 \Lambda_{ {\bf{k}} + {\bf{q}}_{1}/2 }(-{\bf{q}}_{1})
 \langle a^{\dagger}_{ {\bf{k}} + {\bf{q}}_{1}/2 }({\bf{q}}_{1})
 a_{ {\bf{k}} + {\bf{q}}_{1}/2 }({\bf{q}}_{1}) \rangle
\end{equation}
 Assume the $ n_{F}({\bf{k}}) $ are all evaluated at zero temp. as in the
 previous sections.
 Now, taken at face value, one is obliged to compute the thermodynamic
 expectation values of the Bose occupation probabilities assuming the 
 chemical potential for the Bosons is zero (this means that we are allowed
 to create and destroy any number of Bosons). Now such a calculation
 yields an infinite answer for 
 $ \langle c^{\dagger}_{ {\bf{k}} }c_{ {\bf{k}} } \rangle $ as the 
 sum over all $ {\bf{q}}_{1} $ diverges (is proportional to the 
 total number of fermions). This is the reason why a more nefarious
 approach may be necessary in dealing with fermions at finite temperature.
 For this the only guide is that all the finite temp. dynamical correlation
 functions involving the number conserving object 
 $ c^{\dagger}_{ {\bf{k+q/2}} } c_{ {\bf{k-q/2}} } $ should be correctly
 reproduced. Hopefully the commutation rules involving these number
 conserving products are not damaged in the bargain.
 For this let us start with the simplest case namely,
 $ \langle c^{\dagger}_{ {\bf{k+q/2}} } c_{ {\bf{k-q/2}} } \rangle $ .
 We know what the answer should be, that is,
\begin{equation}
 \langle c^{\dagger}_{ {\bf{k+q/2}} } c_{ {\bf{k-q/2}} } \rangle
 = \delta_{ {\bf{q = 0}} } n_{F, \beta}({\bf{k}})
\end{equation}
where,
\begin{equation}
 n_{F, \beta}({\bf{k}}) = [ exp(\beta(\epsilon_{ {\bf{k}} }-\mu)) + 1 ]^{-1}
\end{equation}
For this to happen we have to do the following,
\newline
(1) Treat the bosons as before assuming that they are always at
 zero temperature and with zero chemical potential.
 \newline
(2) Assume that all finite temperature effects are lumped into
 the coefficients.
\newline
 (3) Fix the coefficients as before by demanding that the finite temp.
 dynamical moments of the number-conserving products come out right.
\newline
For the dynamical four-point function to come out right we must ensure that,
\begin{equation}
\Lambda^{\beta}_{ {\bf{k}} }({\bf{q}}) = 
 \sqrt{ n_{F, \beta}({\bf{k+q/2}})(1 - n_{F, \beta}({\bf{k-q/2}})) } 
\end{equation}
 Unfortunately, the coefficient 
 $ \Lambda^{\beta}_{ {\bf{k}} }({\bf{q}}) $ is no
 longer either zero or one. This renders matters even more difficult.
 But the dynamical six-point function is still the same and has to be correctly
 given as in the zero-temp case.
\[
I = \langle c^{\dagger}_{ {\bf{k+q/2}} }c_{ {\bf{k-q/2}} }
c^{\dagger}_{ {\bf{k^{'}+q^{'}/2}} }c_{ {\bf{k^{'}-q^{'}/2}} }
c^{\dagger}_{ {\bf{k^{''}+q^{''}/2}} }c_{ {\bf{k^{''}-q^{''}/2}} } \rangle
\]
\[
 = [(1 - n_{F, \beta}({\bf{k-q/2}}))(1 - n_{F, \beta}({\bf{k^{'}-q^{'}/2}}))
n_{F, \beta}({\bf{k+q/2}})\delta_{ {\bf{k+q/2}}, {\bf{k^{''} - q^{''}/2}} }
\delta_{ {\bf{k-q/2}}, {\bf{k^{'} + q^{'}/2}} }
\delta_{ {\bf{k^{'}-q^{'}/2}}, {\bf{k^{''} + q^{''}/2}} }
\]
\begin{equation}
 - (1 - n_{F, \beta}({\bf{k-q/2}}))n_{F, \beta}({\bf{k^{'}+q^{'}/2}})
n_{F, \beta}({\bf{k+q/2}})
\delta_{ {\bf{k-q/2}}, {\bf{k^{''} + q^{''}/2}} }
\delta_{ {\bf{k+q/2}}, {\bf{k^{'} - q^{'}/2}} }
\delta_{ {\bf{k^{'}+q^{'}/2}}, {\bf{k^{''} - q^{''}/2}} } ]
\end{equation}
In terms of the Bose fields we have,
\begin{equation}
I = \Lambda^{\beta}_{ {\bf{k}} }({\bf{q}})
\Lambda^{\beta}_{ {\bf{k}}^{''} }(-{\bf{q}}^{''})
 \Gamma_{ {\bf{k}}, {\bf{k}}^{''} }^{ -{\bf{q}}, {\bf{q}}^{''} }
({\bf{k}}^{'}, {\bf{q}}^{'})
\end{equation}
Equating these two we arrive at a formula for the coefficient $ \Gamma $.
 It is somewhat different from the zero temp case. But we can remedy
 that by choosing,
\[
\Gamma_{ {\bf{k}}_{1}, {\bf{k}}_{2} }^{ {\bf{q}}_{1}, {\bf{q}}_{2} }
({\bf{k}}, {\bf{q}}) = 
F(\Lambda^{\beta}_{ {\bf{k}}_{1} }(-{\bf{q}}_{1})
\Lambda^{\beta}_{ {\bf{k}}_{2} }(-{\bf{q}}_{2}))
[\sqrt{ (1 - n_{F, \beta}({\bf{k}}_{1}+{\bf{q}}_{1}/2))
 (1 - n_{F, \beta}({\bf{k}}_{2}+{\bf{q}}_{2}/2)) }
\]
\[
\delta_{ {\bf{k_{1}-q_{1}/2}}, {\bf{k_{2}-q_{2}/2}} }
\delta_{ {\bf{k_{1}+q_{1}/2}}, {\bf{k+q/2}} }
\delta_{ {\bf{k-q/2}}, {\bf{k_{2}+q_{2}/2}} }
\]
\[
 - \sqrt{ n_{F, \beta}({\bf{k}}_{1}-{\bf{q}}_{1}/2)
n_{F, \beta}({\bf{k}}_{2}-{\bf{q}}_{2}/2) }
\]
\begin{equation}
\delta_{ {\bf{k_{1}+q_{1}/2}}, {\bf{k_{2}+q_{2}/2}} }
\delta_{ {\bf{k_{1}-q_{1}/2}}, {\bf{k-q/2}} }
\delta_{ {\bf{k+q/2}}, {\bf{k_{2}-q_{2}/2}} } ]
\end{equation}
and,
\begin{equation}
F(x) = 1\mbox{   }if\mbox{   }x \neq 0\mbox{   }
and\mbox{   }F(x) = 0\mbox{   }for\mbox{   }x = 0
\end{equation}
For any temperature above zero $ F(x) = 1 $ always. But for exactly
 at zero temp, the original case is recovered. But more importantly, all
 the right sort of dynamical correlation functions are recovered if we
 choose in addition, the following formula for the kinetic energy
 operator,
\begin{equation}
K = \sum_{ {\bf{k}}, {\bf{q}} }\omega_{ {\bf{k}} }({\bf{q}})
a^{\dagger}_{ {\bf{k}} }({\bf{q}})a_{ {\bf{k}} }({\bf{q}})
\end{equation}
and, 
\begin{equation}
 \omega_{ {\bf{k}} }({\bf{q}}) = F(\Lambda^{\beta}_{ {\bf{k}} }(-{\bf{q}}))
(\frac{ {\bf{k.q}} }{m})
\end{equation}
These choices ensure that all the finite temp. dynamical moments of
 $ c^{\dagger}_{ {\bf{k+q/2}} }c_{ {\bf{k-q/2}} } $ are correctly
 recovered. 
\subsection{ Demonstration of Emergence of Superconductivity }
 Consider two interactions, one the coulomb repulsion and the
 other electron-phonon interaction :
\begin{equation}
H_{I} = \sum_{ {\bf{q}} \neq 0 }\frac{ v_{ {\bf{q}} } }{2V}
\sum_{ {\bf{k}}, {\bf{k}}^{'} }
[\Lambda^{\beta}_{ {\bf{k}} }({\bf{q}})a_{ {\bf{k}} }(-{\bf{q}})
 + \Lambda^{\beta}_{ {\bf{k}} }(-{\bf{q}})a^{\dagger}_{ {\bf{k}} }({\bf{q}})]
[\Lambda^{\beta}_{ {\bf{k}}^{'} }(-{\bf{q}})a_{ {\bf{k}}^{'} }({\bf{q}})
 + \Lambda^{\beta}_{ {\bf{k}}^{'} }({\bf{q}})
a^{\dagger}_{ {\bf{k}}^{'} }(-{\bf{q}})]
\end{equation}
\begin{equation}
H_{e-phon} = \sum_{ {\bf{q}} \neq 0 }\frac{ M_{ {\bf{q}} } }{ V^{\frac{1}{2}} }
\sum_{ {\bf{k}} }
[\Lambda^{\beta}_{ {\bf{k}} }({\bf{q}})a_{ {\bf{k}} }(-{\bf{q}})
 + \Lambda^{\beta}_{ {\bf{k}} }(-{\bf{q}})a^{\dagger}_{ {\bf{k}} }({\bf{q}})]
[b_{ {\bf{q}} } + b^{\dagger}_{ -{\bf{q}} }]
\end{equation}
The displacement operators are,
\begin{equation}
P_{ {\bf{q}} } = b_{ {\bf{q}} } + b^{\dagger}_{ -{\bf{q}} }
\end{equation}
and,
\begin{equation}
X_{ {\bf{q}} } = (\frac{i}{2})
 (b_{ -{\bf{q}} } - b^{\dagger}_{ {\bf{q}} })
\end{equation}
The equations of motion for the Bose fields are,
\[
i \mbox{ }\frac{\partial}{ \partial t }a^{t}_{ {\bf{k}} }({\bf{q}})
 = \omega_{ {\bf{k}} }({\bf{q}})a^{t}_{ {\bf{k}} }({\bf{q}})
 + (\frac{ v_{ {\bf{q}} } }{V})
\Lambda^{\beta}_{ {\bf{k}} }(-{\bf{q}})\sum_{ {\bf{k}}^{'} }[
\Lambda^{\beta}_{ {\bf{k}}^{'} }(-{\bf{q}})a^{t}_{ {\bf{k}}^{'} }({\bf{q}})
+ \Lambda^{\beta}_{ {\bf{k}}^{'} }({\bf{q}})
a^{\dagger t}_{ {\bf{k}}^{'} }(-{\bf{q}}) ]
\]
\begin{equation}
 + \frac{ M_{ {\bf{q}} } }{ V^{\frac{1}{2}} }
\Lambda^{\beta}_{ {\bf{k}} }(-{\bf{q}})
 [ b_{ {\bf{q}} } + b^{\dagger}_{ -{\bf{q}} } ]
\end{equation}
\begin{equation}
i\frac{ \partial }{\partial t}X_{ {\bf{q}} }
 = \frac{ i\mbox{ }M_{ {\bf{q}} } }{ V^{\frac{1}{2}} }
\sum_{ {\bf{k}}^{'} }
[\Lambda_{ {\bf{k}}^{'} }({\bf{q}})a_{ {\bf{k}}^{'} }(-{\bf{q}})
+ \Lambda_{ {\bf{k}}^{'} }(-{\bf{q}})a^{\dagger}_{ {\bf{k}}^{'} }({\bf{q}}) ]
+ \frac{ i\Omega_{ {\bf{q}} } }{2}P_{ -{\bf{q}} }
\end{equation}
\begin{equation}
i\frac{ \partial }{\partial t}P_{ {\bf{q}} }
 = \frac{ 2\mbox{ }\Omega_{ {\bf{q}} } }{i}X_{ -{\bf{q}} }
\end{equation}

\[
i \mbox{ }\frac{\partial}{ \partial t }a^{\dagger t}_{ {\bf{k}} }(-{\bf{q}})
 = -\omega_{ {\bf{k}} }(-{\bf{q}})a^{\dagger t}_{ {\bf{k}} }(-{\bf{q}})
 - (\frac{ v_{ {\bf{q}} } }{V})
\Lambda^{\beta}_{ {\bf{k}} }({\bf{q}})\sum_{ {\bf{k}}^{'} }[
\Lambda^{\beta}_{ {\bf{k}}^{'} }(-{\bf{q}})a^{t}_{ {\bf{k}}^{'} }({\bf{q}})
+ \Lambda^{\beta}_{ {\bf{k}}^{'} }({\bf{q}})a^{\dagger t}
_{ {\bf{k}}^{'} }(-{\bf{q}}) ]
\]
\begin{equation}
 - \frac{ M_{ {\bf{q}} } }{ V^{\frac{1}{2}} }
\Lambda^{\beta}_{ {\bf{k}} }({\bf{q}})
 [ b_{ {\bf{q}} } + b^{\dagger}_{ -{\bf{q}} } ]
\end{equation}

\[
(i\frac{\partial}{\partial t} - \omega_{ {\bf{k}} }({\bf{q}}))
\frac{ -i\langle T a^{t}_{ {\bf{k}} }({\bf{q}})
a^{\dagger}_{ {\bf{k}}^{'} }({\bf{q}}^{'}) \rangle }{\langle T 1 \rangle}
 = \delta_{ {\bf{k}}, {\bf{k}}^{'} }
\delta_{ {\bf{q}}, {\bf{q}}^{'} }\delta(t)
\]
\[
+ (\frac{ v_{ {\bf{q}} } }{V})
\Lambda^{\beta}_{ {\bf{k}} }(-{\bf{q}})
\sum_{ {\bf{k}}^{''} }[\Lambda^{\beta}_{ {\bf{k}}^{''} }(-{\bf{q}})
\frac{ -i\langle T \mbox{ }a^{t}_{ {\bf{k}}^{''} }({\bf{q}})
a^{\dagger}_{ {\bf{k}}^{'} }({\bf{q}}^{'}) \rangle }{\langle T 1 \rangle}
+ \Lambda^{\beta}_{ {\bf{k}}^{''} }({\bf{q}})
\frac{ -i\langle T \mbox{ }a^{\dagger t}_{ {\bf{k}}^{''} }(-{\bf{q}})
a^{\dagger}_{ {\bf{k}}^{'} }({\bf{q}}^{'}) \rangle }{\langle T 1 \rangle}]
\]
\begin{equation}
+ \frac{ M_{ {\bf{q}} } }{ V^{\frac{1}{2}} }
\Lambda^{\beta}_{ {\bf{k}} }(-{\bf{q}})
 \frac{ -i \langle T \mbox{ }P^{t}_{ {\bf{q}} }
a^{\dagger}_{ {\bf{k}}^{'} }({\bf{q}}^{'}) \rangle }
{\langle T1 \rangle}
\end{equation}

\[
(i\frac{\partial}{\partial t} + \omega_{ {\bf{k}} }(-{\bf{q}}))
\frac{ -i\langle T \mbox{ }a^{\dagger t}_{ {\bf{k}} }(-{\bf{q}})
a^{\dagger}_{ {\bf{k}}^{'} }({\bf{q}}^{'}) \rangle }{\langle T 1 \rangle}
 = 
\]
\[
- (\frac{ v_{ {\bf{q}} } }{V})
\Lambda^{\beta}_{ {\bf{k}} }({\bf{q}})
\sum_{ {\bf{k}}^{''} }[\Lambda^{\beta}_{ {\bf{k}}^{''} }(-{\bf{q}})
\frac{ -i\langle T \mbox{ }a^{t}_{ {\bf{k}}^{''} }({\bf{q}})
a^{\dagger}_{ {\bf{k}}^{'} }({\bf{q}}^{'}) \rangle }{\langle T 1 \rangle}
+ \Lambda^{\beta}_{ {\bf{k}}^{''} }({\bf{q}})
\frac{ -i\langle T \mbox{ }a^{\dagger t}_{ {\bf{k}}^{''} }(-{\bf{q}})
a^{\dagger}_{ {\bf{k}}^{'} }({\bf{q}}^{'}) \rangle }{\langle T 1 \rangle}]
\]
\begin{equation}
- \frac{ M_{ {\bf{q}} } }{ V^{\frac{1}{2}} }
\Lambda^{\beta}_{ {\bf{k}} }({\bf{q}})
 \frac{ -i \langle T \mbox{ }P^{t}_{ {\bf{q}} } \mbox{  }
a^{\dagger}_{ {\bf{k}}^{'} }({\bf{q}}^{'}) \rangle }
{\langle T1 \rangle}
\end{equation}

The corresponding propagators with phonons are,
\begin{equation}
i \frac{ \partial }{ \partial t }
\frac{ -i \langle T \mbox{ }P^{t}_{ {\bf{q}} }\mbox{  }
a^{\dagger}_{ {\bf{k}}^{'} }({\bf{q}}^{'})
 \rangle }{ \langle T 1 \rangle }
 = (\frac{ 2 \Omega_{ {\bf{q}} } }{i})
\frac{ -i \langle T \mbox{ }X^{t}_{ -{\bf{q}} }
\mbox{  }a^{\dagger}_{ {\bf{k}}^{'} }({\bf{q}}^{'})
 \rangle }{ \langle T 1 \rangle }
\end{equation}
\[
i \frac{ \partial }{ \partial t }
\frac{ -i \langle T \mbox{ }X^{t}_{ -{\bf{q}} }
\mbox{  }a^{\dagger}_{ {\bf{k}}^{'} }({\bf{q}}^{'})
 \rangle }{ \langle T 1 \rangle }
 = (\frac{ i\mbox{ }M_{ -{\bf{q}} } }{ V^{\frac{1}{2}} })
\sum_{ {\bf{k}}^{''} }[ \Lambda^{\beta}_{ {\bf{k}}^{''} }(-{\bf{q}})
\frac{ -i \langle T \mbox{ }a^{t}_{ {\bf{k}}^{''} }({\bf{q}})
a^{\dagger}_{ {\bf{k}}^{'} }({\bf{q}}^{'})
 \rangle }{ \langle T 1 \rangle }
 + \Lambda^{\beta}_{ {\bf{k}}^{''} }({\bf{q}})
\frac{ -i \langle T \mbox{ }a^{\dagger t}_{ {\bf{k}}^{''} }(-{\bf{q}})
a^{\dagger}_{ {\bf{k}}^{'} }({\bf{q}}^{'})
 \rangle }{ \langle T 1 \rangle } ]
\]
\begin{equation}
 + (\frac{ i \mbox{ }\Omega_{ {\bf{q}} } }{2})
\frac{ -i\langle T \mbox{ }P^{t}_{ {\bf{q}} }
\mbox{   }a^{\dagger}_{ {\bf{k}}^{'} }({\bf{q}}^{'}) \rangle }
{ \langle T 1 \rangle }
\end{equation}
Transform to frequency domain and we have the corresponding equations :
\[
G_{1}( {\bf{q}}, {\bf{k}}^{'}, {\bf{q}}^{'}; n) =
\Lambda_{ {\bf{k}}^{'} }(-{\bf{q}})
\frac{ \delta_{ {\bf{q}}, {\bf{q}}^{'} } }
{-i\mbox{ } \beta( i\omega_{n} - \omega_{ {\bf{k}}^{'} }({\bf{q}}) ) }
\]
\begin{equation}
+ f_{n}({\bf{q}})[G_{1}( {\bf{q}}, {\bf{k}}^{'}, {\bf{q}}^{'}; n)
+ G_{2}( {\bf{q}}, {\bf{k}}^{'}, {\bf{q}}^{'}; n) ]
+ g_{n}({\bf{q}})G_{3}( {\bf{q}}, {\bf{k}}^{'}, {\bf{q}}^{'}; n)  
\end{equation}
\begin{equation}
G_{2}( {\bf{q}}, {\bf{k}}^{'}, {\bf{q}}^{'}; n)
 = f^{*}_{n}(-{\bf{q}})
[G_{1}( {\bf{q}}, {\bf{k}}^{'}, {\bf{q}}^{'}; n) 
 + G_{2}( {\bf{q}}, {\bf{k}}^{'}, {\bf{q}}^{'}; n)]
+ g^{*}_{n}(-{\bf{q}})G_{3}( {\bf{q}}, {\bf{k}}^{'}, {\bf{q}}^{'}; n)
\end{equation}
\begin{equation}
i\mbox{ }\omega_{n}\mbox{ }G_{3}( {\bf{q}}, {\bf{k}}^{'}, {\bf{q}}^{'}; n)
 = (\frac{2 \mbox{ }\Omega_{ {\bf{q}} } }{i})
G_{4}( {\bf{q}}, {\bf{k}}^{'}, {\bf{q}}^{'}; n)
\end{equation}
\begin{equation}
i\mbox{ }\omega_{n}\mbox{ }G_{4}( {\bf{q}}, {\bf{k}}^{'}, {\bf{q}}^{'}; n)
 = \mbox{ }i\mbox{ }[G_{1}( {\bf{q}}, {\bf{k}}^{'}, {\bf{q}}^{'}; n)
 + G_{2}( {\bf{q}}, {\bf{k}}^{'}, {\bf{q}}^{'}; n) ]
 + (\frac{ i \mbox{  }\Omega_{ {\bf{q}} } }{2})
 G_{3}( {\bf{q}}, {\bf{k}}^{'}, {\bf{q}}^{'}; n)
\end{equation}
The solutions may be written down as,
\begin{equation}
G_{3}({\bf{q}}, {\bf{k}}^{'}, {\bf{q}}^{'}; n)
 = -(\frac{ 2 \mbox{  }\Omega_{ {\bf{q}} } }
{ \Omega^{2}_{ {\bf{q}} } + \omega_{n}^{2} })
[G_{1}( {\bf{q}}, {\bf{k}}^{'}, {\bf{q}}^{'}; n)
 + G_{2}( {\bf{q}}, {\bf{k}}^{'}, {\bf{q}}^{'}; n)]
\end{equation}
\begin{equation}
G_{2}({\bf{q}}, {\bf{k}}^{'}, {\bf{q}}^{'}; n)
 = (\frac{ \Gamma_{n}({\bf{q}}) }{1 - \Gamma_{n}({\bf{q}})})
G_{1}({\bf{q}}, {\bf{k}}^{'}, {\bf{q}}^{'}; n)
\end{equation}
\begin{equation}
\Gamma_{n}({\bf{q}}) = f^{*}_{n}(-{\bf{q}})
 - \frac{ 2 \mbox{  }\Omega_{ {\bf{q}} } }
{ \Omega^{2}_{ {\bf{q}} } + \omega_{n}^{2} }
g^{*}_{n}(-{\bf{q}})
\end{equation}
and finally,
\begin{equation}
G_{1}({\bf{q}}, {\bf{k}}^{'}, {\bf{q}}^{'}; n)
 = \delta_{ {\bf{q}}, {\bf{q}}^{'} }
\frac{ F_{n}({\bf{q}}) \Lambda_{ {\bf{k}}^{'} }(-{\bf{q}}) }
{-i\mbox{ } \beta(i \mbox{  }\omega_{n} - \omega_{ {\bf{k}}^{'} }({\bf{q}}) ) }
\end{equation}
and,
\begin{equation}
F_{n}({\bf{q}})
 = [ 1 - \frac{ f_{n}({\bf{q}}) }{ 1 - \Gamma_{n}({\bf{q}}) }
 + (\frac{ 2 \mbox{  }\Omega_{ {\bf{q}} } }
{ \Omega^{2}_{ {\bf{q}} } + \omega^{2}_{n} })
(\frac{ g_{n}({\bf{q}}) }{1 - \Gamma_{n}({\bf{q}}) }) ]^{-1}
\end{equation}
here,
\begin{equation}
f_{n}({\bf{q}}) = (\frac{ v_{ {\bf{q}} } }{V})
\sum_{ {\bf{k}} }
\frac{ (  \Lambda^{\beta}_{ {\bf{k}} }(-{\bf{q}})  )^{2} }
{ ( i\omega_{n} - \omega_{ {\bf{k}} }({\bf{q}}) ) }
\end{equation}
\begin{equation}
g_{n}({\bf{q}}) = (\frac{ |M_{ {\bf{q}} }|^{2} }{ V })
\sum_{ {\bf{k}} }
\frac{ (  \Lambda^{\beta}_{ {\bf{k}} }(-{\bf{q}})  )^{2} }
{ ( i\omega_{n} - \omega_{ {\bf{k}} }({\bf{q}}) ) }
\end{equation}
For the four-point function we have to proceed as follows, the quantity
 of interest is(with phonons and coulomb),
\[
\frac{1}{i^{2}}\langle T (a^{t_{1}}_{ {\bf{k}}_{1} }({\bf{q}}_{1})
 \mbox{ }a^{t_{2}}_{ {\bf{k}}_{2} }({\bf{q}}_{2})
a^{\dagger t^{'}_{2}}_{ {\bf{k}}^{'}_{2} }({\bf{q}}^{'}_{2})
a^{\dagger t^{'}_{1}}_{ {\bf{k}}^{'}_{1} }({\bf{q}}^{'}_{1})) \rangle
\]
for this we as usual decompose as follows,
\[
\frac{1}{i^{2}}\langle T (a^{t_{1}}_{ {\bf{k}}_{1} }({\bf{q}}_{1})
 \mbox{ }a^{t_{2}}_{ {\bf{k}}_{2} }({\bf{q}}_{2})
a^{\dagger t^{'}_{2}}_{ {\bf{k}}^{'}_{2} }({\bf{q}}^{'}_{2})
a^{\dagger t^{'}_{1}}_{ {\bf{k}}^{'}_{1} }({\bf{q}}^{'}_{1})) \rangle
\]
\begin{equation}
 = \sum_{ n }\mbox{ }exp(\omega_{n}\mbox{ }t_{1})
\frac{1}{i^{2}}\langle T (a^{n}_{ {\bf{k}}_{1} }({\bf{q}}_{1})
 \mbox{ }a^{t_{2}}_{ {\bf{k}}_{2} }({\bf{q}}_{2})
a^{\dagger t^{'}_{2}}_{ {\bf{k}}^{'}_{2} }({\bf{q}}^{'}_{2})
a^{\dagger t^{'}_{1}}_{ {\bf{k}}^{'}_{1} }({\bf{q}}^{'}_{1})) \rangle
\end{equation}
\[
(i \mbox{ }\omega_{n} - \omega_{ {\bf{k}}_{1} }({\bf{q}}_{1}) )
\frac{1}{i^{2}}
\langle T(a^{n}_{ {\bf{k}}_{1} }({\bf{q}}_{1})
a^{t_{2}}_{ {\bf{k}}_{2} }({\bf{q}}_{2})
a^{\dagger t^{'}_{2}}_{ {\bf{k}}^{'}_{2} }({\bf{q}}^{'}_{2})
a^{\dagger t^{'}_{1}}_{ {\bf{k}}^{'}_{1} }({\bf{q}}^{'}_{1}) )
\rangle
\]
\[
 = F_{1}({\bf{k}}_{1}\mbox{ }{\bf{q}}_{1}\mbox{ }n,
 {\bf{k}}_{2}\mbox{ }{\bf{q}}_{2}\mbox{ }t_{2},
 {\bf{k}}^{'}_{2}\mbox{ }{\bf{q}}^{'}_{2}\mbox{ }t^{'}_{2},
 {\bf{k}}^{'}_{1}\mbox{ }{\bf{q}}^{'}_{1}\mbox{ }t^{'}_{1})
\]
\[
+ (\frac{ \Lambda_{ {\bf{k}}_{1} }(-{\bf{q}}_{1}) }{V})
[v_{ {\bf{q}}_{1} } - (\frac{ 2 \mbox{ }\Omega_{ {\bf{q}}_{1} } }
{ \omega^{2}_{n} + \Omega^{2}_{ {\bf{q}}_{1} } })
|M_{ {\bf{q}}_{1} }|^{2}]
\]
\begin{equation}
\times [{\tilde{F}}_{1}({\bf{q}}_{1}n,
{\bf{k}}_{2}\mbox{ }{\bf{q}}_{2}\mbox{ }t_{2},
{\bf{k}}^{'}_{2}\mbox{ }{\bf{q}}^{'}_{2}\mbox{ }t^{'}_{2},
{\bf{k}}^{'}_{1}\mbox{ }{\bf{q}}^{'}_{1}\mbox{ }t^{'}_{1})
+ {\tilde{F}}_{2}({\bf{q}}_{1}n,
{\bf{k}}_{2}\mbox{ }{\bf{q}}_{2}\mbox{ }t_{2},
{\bf{k}}^{'}_{2}\mbox{ }{\bf{q}}^{'}_{2}\mbox{ }t^{'}_{2},
{\bf{k}}^{'}_{1}\mbox{ }{\bf{q}}^{'}_{1}\mbox{ }t^{'}_{1}) ]/
(1 - \Gamma_{n}({\bf{q}}_{1}) - \Gamma^{*}_{n}(-{\bf{q}}_{1}))
\end{equation}
\newline
\[
{\tilde{F}}_{1}({\bf{q}}_{1}n,
{\bf{k}}_{2}\mbox{ }{\bf{q}}_{2}\mbox{ }t_{2},
{\bf{k}}^{'}_{2}\mbox{ }{\bf{q}}^{'}_{2}\mbox{ }t^{'}_{2},
{\bf{k}}^{'}_{1}\mbox{ }{\bf{q}}^{'}_{1}\mbox{ }t^{'}_{1})
\]
\begin{equation}
 = \sum_{ {\bf{k}}_{1} }
\frac{ \Lambda_{ {\bf{k}}_{1} }(-{\bf{q}}_{1}) }{
( i\omega_{n} - \omega_{ {\bf{k}}_{1} }({\bf{q}}_{1}) ) }
F_{1}({\bf{k}}_{1}{\bf{q}}_{1}n,
{\bf{k}}_{2}\mbox{ }{\bf{q}}_{2}\mbox{ }t_{2},
{\bf{k}}^{'}_{2}\mbox{ }{\bf{q}}^{'}_{2}\mbox{ }t^{'}_{2},
{\bf{k}}^{'}_{1}\mbox{ }{\bf{q}}^{'}_{1}\mbox{ }t^{'}_{1})
\end{equation}
\[
{\tilde{F}}_{2}({\bf{q}}_{1}n,
{\bf{k}}_{2}\mbox{ }{\bf{q}}_{2}\mbox{ }t_{2},
{\bf{k}}^{'}_{2}\mbox{ }{\bf{q}}^{'}_{2}\mbox{ }t^{'}_{2},
{\bf{k}}^{'}_{1}\mbox{ }{\bf{q}}^{'}_{1}\mbox{ }t^{'}_{1})
\]
\begin{equation}
 = \sum_{ {\bf{k}}_{1} }
\frac{ \Lambda_{ {\bf{k}}_{1} }({\bf{q}}_{1}) }{
( i\omega_{n} + \omega_{ {\bf{k}}_{1} }(-{\bf{q}}_{1}) ) }
F_{2}({\bf{k}}_{1}{\bf{q}}_{1}n,
{\bf{k}}_{2}\mbox{ }{\bf{q}}_{2}\mbox{ }t_{2},
{\bf{k}}^{'}_{2}\mbox{ }{\bf{q}}^{'}_{2}\mbox{ }t^{'}_{2},
{\bf{k}}^{'}_{1}\mbox{ }{\bf{q}}^{'}_{1}\mbox{ }t^{'}_{1})
\end{equation}

\subsection{ Bosons at zero temperature }

 In this subsection we do what we did earlier except that here we
 are bosonizing the bosons. Let $ b_{ {\bf{k}} } $
 and  $ b^{\dagger}_{ {\bf{k}} } $ be the Bose fields in question. 
 Analogous to Fermi sea displacements we introduce Bose 
 condensate displacements. 
\[
b^{\dagger}_{ {\bf{k+q/2}} } b_{ {\bf{k-q/2}} }
 = N \mbox{  }\delta_{ {\bf{k}}, 0 } \delta_{ {\bf{q}}, 0 }
 + (\sqrt{N})
[\delta_{ {\bf{k+q/2}}, 0 }\mbox{  }d_{ {\bf{k}} }(-{\bf{q}})
 + \delta_{ {\bf{k-q/2}}, 0 }\mbox{  }d^{\dagger}_{ {\bf{k}} }({\bf{q}})]
\]
\[
 + \sum_{ {\bf{q}}_{1} }T_{1}({\bf{k}}, {\bf{q}}, {\bf{q}}_{1})
 d^{\dagger}_{ {\bf{k+q/2-q_{1}/2}} }({\bf{q}}_{1})
 d_{ {\bf{k-q_{1}/2}} }(-{\bf{q}}+{\bf{q}}_{1})
\]
\begin{equation}
 +  \sum_{ {\bf{q}}_{1} }T_{2}({\bf{k}}, {\bf{q}}, {\bf{q}}_{1})
 d^{\dagger}_{ {\bf{k-q/2+q_{1}/2}} }({\bf{q}}_{1})
 d_{ {\bf{k+q_{1}/2}} }(-{\bf{q}}+{\bf{q}}_{1})
\end{equation}
 To make all the dynamical moments of 
 $ b^{\dagger}_{ {\bf{k+q/2}} } b_{ {\bf{k-q/2}} } $ come out right and the
 commutation rules amongst them also come out right provided we choose
 $ T_{1} $ and $ T_{2} $ such that,
\[
(\sqrt{N})^{2}\delta_{ {\bf{k+q/2}}, 0 }
\delta_{ {\bf{k^{''}-q^{''}/2}}, 0 }
\sum_{ {\bf{q}}_{1} }T_{1}({\bf{k}}^{'}, {\bf{q}}^{'}, {\bf{q}}_{1})
\langle d_{ {\bf{k}} }(-{\bf{q}})
d^{\dagger}_{ {\bf{k}}^{'} + {\bf{q}}^{'}/2 - {\bf{q}}_{1}/2 }
({\bf{q}}_{1})d_{ {\bf{k}}^{'} - {\bf{q}}_{1}/2 }
(-{\bf{q}}^{'} + {\bf{q}}_{1}) d^{\dagger}_{ {\bf{k}}^{''} }({\bf{q}}^{''})
 \rangle
\]
\[
+(\sqrt{N})^{2}\delta_{ {\bf{k+q/2}}, 0 }
\delta_{ {\bf{k^{''}-q^{''}/2}}, 0 }
\sum_{ {\bf{q}}_{1} }T_{2}({\bf{k}}^{'}, {\bf{q}}^{'}, {\bf{q}}_{1})
\langle d_{ {\bf{k}} }(-{\bf{q}})
d^{\dagger}_{ {\bf{k}}^{'} - {\bf{q}}^{'}/2 + {\bf{q}}_{1}/2 }
({\bf{q}}_{1})d_{ {\bf{k}}^{'} + {\bf{q}}_{1}/2 }
(-{\bf{q}}^{'} + {\bf{q}}_{1}) d^{\dagger}_{ {\bf{k}}^{''} }({\bf{q}}^{''})
 \rangle
\]
 = 
\begin{equation}
N\mbox{ }\delta_{ {\bf{k+q/2}}, 0 }
\delta_{ {\bf{k+q/2}}, {\bf{k^{''}-q^{''}/2}} }
\delta_{ {\bf{k-q/2}}, {\bf{k^{'}+q^{'}/2}} }
\delta_{ {\bf{k^{'}-q^{'}/2}}, {\bf{k^{''}+q^{''}/2}} }
\end{equation}
This means,
\begin{equation}
T_{1}( -{\bf{q}}-{\bf{q}}^{'}/2, {\bf{q}}^{'}, -{\bf{q}} )
 = 1; \mbox{   }{\bf{q}} \neq 0; \mbox{   }{\bf{q}}^{'} \neq 0
\end{equation}
\begin{equation}
T_{2}( -{\bf{q}}^{'}/2, {\bf{q}}^{'}, -{\bf{q}} )
 = 0; \mbox{   }{\bf{q}} \neq 0; \mbox{   }{\bf{q}}^{'} \neq 0
\end{equation}
in order for the kinetic energy operator to have the form,
\begin{equation}
K = \sum_{ {\bf{k}} }\epsilon_{ {\bf{k}} } \mbox{  }
d^{\dagger}_{ (1/2){\bf{k}} }({\bf{k}})
\mbox{ }d_{ (1/2){\bf{k}} }({\bf{k}})
\end{equation}
we must have,
\begin{equation}
\epsilon_{ {\bf{k+q_{1}/2}} }T_{1}( {\bf{k+q_{1}/2}}, {\bf{0}}, {\bf{q}}_{1})
 + \epsilon_{ {\bf{k-q_{1}/2}} }T_{2}( {\bf{k-q_{1}/2}}, {\bf{0}}, {\bf{q}}_{1})
 = \delta_{ {\bf{k}}, {\bf{q}}_{1}/2 }\epsilon_{ {\bf{q}}_{1} }
\label{EQ1}
\end{equation}
 In order for,
\[
  \sum_{ {\bf{k}} }b^{ \dagger }_{ {\bf{k}} }b_{ {\bf{k}} }
 = N
\]
we must have,
\begin{equation}
T_{1}({\bf{k+q_{1}/2}}, {\bf{0}}, {\bf{q_{1}}}) 
 + T_{2}({\bf{k-q_{1}/2}}, {\bf{0}}, {\bf{q_{1}}}) 
 = 0
\label{EQ2}
\end{equation}
 In order for the commutation rules amongst the 
 $ b^{\dagger}_{ {\bf{k+q/2}} } b_{ {\bf{k-q/2}} } $ to come out right,
 we must have in addition to all the above relations,
\begin{equation}
T_{2}({\bf{q}}/2, {\bf{q}}, -{\bf{q}}^{'}) = 0
\end{equation}
\begin{equation}
T_{2}(-{\bf{q}}/2, {\bf{q}}, {\bf{q}}+{\bf{q}}^{'}) = 0
\end{equation}
Any choice of $ T_{1} $ and $ T_{2} $ consistent with the above
 relations should suffice.
From the above relations (Eq.(~\ref{EQ1}) and Eq.(~\ref{EQ2})), 
 we have quite unambiguously,
\begin{equation}
T_{1}({\bf{k+q_{1}/2}}, {\bf{0}}, {\bf{q}}_{1}) = 
\delta_{ {\bf{k}}, {\bf{q}}_{1}/2 }
\end{equation}
\begin{equation}
T_{2}({\bf{k-q_{1}/2}}, {\bf{0}}, {\bf{q}}_{1}) =
-\delta_{ {\bf{k}}, {\bf{q}}_{1}/2 }
\end{equation}
\begin{equation}
T_{1}(-{\bf{q}}-{\bf{q}}^{'}/2, {\bf{q}}^{'}, -{\bf{q}}) = 1
\end{equation}
and,
\begin{equation}
T_{1}({\bf{k}}, {\bf{q}}, {\bf{q}}_{1}) = 0; \mbox{  }otherwise
\end{equation}
\begin{equation}
T_{2}({\bf{k}}, {\bf{q}}, {\bf{q}}_{1}) = 0; \mbox{  }otherwise
\end{equation}
This means that we may rewrite the formula for 
 $ b^{\dagger}_{ {\bf{k+q/2}} }b_{ {\bf{k-q/2}} } $
as follows,
\[
b^{\dagger}_{ {\bf{k+q/2}} }b_{ {\bf{k-q/2}} }
 = N \delta_{ {\bf{k}}, 0 }\delta_{ {\bf{q}}, 0 } +
(\sqrt{N})
[\delta_{ {\bf{k+q/2}}, 0 }d_{ {\bf{k}} }(-{\bf{q}})  
 + \delta_{ {\bf{k-q/2}}, 0 }d^{\dagger}_{ {\bf{k}} }({\bf{q}})]
\]
\[
+ 
d^{\dagger}_{ (1/2){\bf{k+q/2}} }({\bf{k+q/2}})
d_{ (1/2){\bf{k-q/2}} }({\bf{k-q/2}})
\]
\begin{equation}
-\delta_{ {\bf{k}}, 0 }\delta_{ {\bf{q}}, 0 }
\sum_{ {\bf{q}}_{1} }d^{\dagger}_{ {\bf{q}}_{1}/2 }({\bf{q}}_{1})
d_{ {\bf{q}}_{1}/2 }({\bf{q}}_{1})
\end{equation}
Consider an interaction of the type,
\begin{equation}
H_{I} = (\frac{\rho_{0}}{2})
\sum_{ {\bf{q}} \neq 0 }
v_{ {\bf{q}} }\sum_{ {\bf{k}}, {\bf{k}}^{'} }
[\delta_{ {\bf{k+q/2}}, 0 }\mbox{  }d_{ {\bf{k}} }(-{\bf{q}})
 + \delta_{ {\bf{k-q/2}}, 0 }\mbox{  }d^{\dagger}_{ {\bf{k}} }({\bf{q}})]
[\delta_{ {\bf{k^{'}-q/2}}, 0 }\mbox{  }d_{ {\bf{k}}^{'} }({\bf{q}})
 + \delta_{ {\bf{k^{'}+q/2}}, 0 }\mbox{  }
d^{\dagger}_{ {\bf{k}}^{'} }(-{\bf{q}})]
\end{equation}
or,
\begin{equation}
H_{I} = (\frac{\rho_{0}}{2})
\sum_{ {\bf{q}} \neq 0 }
v_{ {\bf{q}} }
[d_{ -{\bf{q}}/2 }(-{\bf{q}})
 + d^{\dagger}_{ {\bf{q}}/2 }({\bf{q}})]
[d_{ {\bf{q}}/2 }({\bf{q}})
 + d^{\dagger}_{ -{\bf{q}}/2 }(-{\bf{q}})]
\end{equation}
and the free case is given by,
\begin{equation}
H_{0} = \sum_{ {\bf{q}} }\epsilon_{ {\bf{q}} }
d^{\dagger}_{ (1/2){\bf{q}} }({\bf{q}})d_{ (1/2){\bf{q}} }({\bf{q}})
\end{equation}
The full hamiltonian may be diagonalised as follows,
\begin{equation}
H = \sum_{ {\bf{q}} }\omega_{ {\bf{q}} }
f^{\dagger}_{ {\bf{q}} }f_{ {\bf{q}} }
\end{equation}
and,
\begin{equation}
f_{ {\bf{q}} } =
 (\frac{ \omega_{ {\bf{q}} } + \epsilon_{ {\bf{q}} }
 + \rho_{0}v_{ {\bf{q}} } }
{ 2 \mbox{ }\omega_{ {\bf{q}} } })^{\frac{1}{2}}
d_{ {\bf{q}}/2 }({\bf{q}})
 + (\frac{ -\omega_{ {\bf{q}} } +
 \epsilon_{ {\bf{q}} } + \rho_{0}v_{ {\bf{q}} } }
{ 2 \mbox{ }\omega_{ {\bf{q}} } })^{\frac{1}{2}}
d^{\dagger}_{ -{\bf{q}}/2 }(-{\bf{q}})
\end{equation}
\begin{equation}
f^{\dagger}_{ -{\bf{q}} } =
(\frac{ -\omega_{ {\bf{q}} } +
 \epsilon_{ {\bf{q}} } + \rho_{0}v_{ {\bf{q}} } }
{ 2 \mbox{ }\omega_{ {\bf{q}} } })^{\frac{1}{2}}
d_{ {\bf{q}}/2 }({\bf{q}})
+
 (\frac{ \omega_{ {\bf{q}} } + \epsilon_{ {\bf{q}} }
 + \rho_{0}v_{ {\bf{q}} } }
{ 2 \mbox{ }\omega_{ {\bf{q}} } })^{\frac{1}{2}}
d^{\dagger}_{ -{\bf{q}}/2 }(-{\bf{q}})
\end{equation}
\begin{equation}
d_{ {\bf{q}}/2 }({\bf{q}})
 = (\frac{ \omega_{ {\bf{q}} } + \epsilon_{ {\bf{q}} }
 + \rho_{0}v_{ {\bf{q}} } }
{ 2 \mbox{ }\omega_{ {\bf{q}} } })^{\frac{1}{2}}f_{ {\bf{q}} }
 - (\frac{ -\omega_{ {\bf{q}} } + \epsilon_{ {\bf{q}} }
 + \rho_{0}v_{ {\bf{q}} } }
{ 2 \mbox{ }\omega_{ {\bf{q}} } })^{\frac{1}{2}}f^{\dagger}_{ -{\bf{q}} } 
\end{equation}
\begin{equation}
d^{\dagger}_{ -{\bf{q}}/2 }(-{\bf{q}})
 = (\frac{ \omega_{ {\bf{q}} } + \epsilon_{ {\bf{q}} }
 + \rho_{0}v_{ {\bf{q}} } }
{ 2 \mbox{ }\omega_{ {\bf{q}} } })^{\frac{1}{2}}f^{\dagger}_{ -{\bf{q}} }
 - (\frac{ -\omega_{ {\bf{q}} } + \epsilon_{ {\bf{q}} }
 + \rho_{0}v_{ {\bf{q}} } }
{ 2 \mbox{ }\omega_{ {\bf{q}} } })^{\frac{1}{2}}f_{ {\bf{q}} }
\end{equation}
\begin{equation}
\omega_{ {\bf{q}} } = \sqrt{ \epsilon^{2}_{ {\bf{q}} } 
+ 2 \rho_{0}v_{ {\bf{q}} }\epsilon_{ {\bf{q}} } }
\end{equation}

From this we may deduce,
\begin{equation}
\langle d^{\dagger}_{ (1/2){\bf{q}} }({\bf{q}})
d_{ (1/2){\bf{q}} }({\bf{q}}) \rangle
 = \frac{ -\omega_{ {\bf{q}} } + \epsilon_{ {\bf{q}} }  
+ \rho_{0}v_{ {\bf{q}} } }{ 2\mbox{ }\omega_{ {\bf{q}} } }
\end{equation}
From this it is possible to write down the filling fraction.
\newline
{\center{ {\bf{FILLING FRACTION}} }}
\begin{equation}
f_{0} = N_{0}/N = 1 - 
(1/N)\sum_{ {\bf{q}} }\langle d^{\dagger}_{ (1/2){\bf{q}} }({\bf{q}})
d_{ (1/2){\bf{q}} }({\bf{q}}) \rangle
\end{equation}
or,
\begin{equation}
f_{0} = N_{0}/N = 1 - (1/2\pi^{2}\rho_{0})\int_{q_{min}}^{\infty}
 \mbox{ }dq\mbox{ }
q^{2} ( \frac{ -\omega_{ {\bf{q}} } + \epsilon_{ {\bf{q}} }
+ \rho_{0}v_{ {\bf{q}} } }{ 2\mbox{ }\omega_{ {\bf{q}} } } )
\end{equation}
Here the lower limit is necessary to ensure that the state 
\[
d_{ (1/2){\bf{q}} }({\bf{q}}) |G\rangle 
\]
has positive norm.
and the value of $ q_{min} $ is given by,
\begin{equation}
-\omega_{ q_{min} } + \epsilon_{ q_{min} }+ \rho_{0}v_{ q_{min} }  = 0
\end{equation}
First define,
\begin{equation}
A_{ {\bf{q}} } = (\frac{ \omega_{ {\bf{q}} } + \epsilon_{ {\bf{q}} }
+ \rho_{0}v_{ {\bf{q}} } }{2 \mbox{ }\omega_{ {\bf{q}} } })^{\frac{1}{2}}
\end{equation}
\begin{equation}
B_{ {\bf{q}} } = (\frac{ -\omega_{ {\bf{q}} } + \epsilon_{ {\bf{q}} }
+ \rho_{0}v_{ {\bf{q}} } }{2 \mbox{ }\omega_{ {\bf{q}} } })^{\frac{1}{2}}
\end{equation}
The density operator is,
\begin{equation}
\rho_{ {\bf{q}} }(t)
  = \sqrt{N}[d_{ -(1/2){\bf{q}} }(-{\bf{q}})(t) +
d^{\dagger}_{ (1/2){\bf{q}} }({\bf{q}})(t)]
+ \sum_{ {\bf{k}} }d^{\dagger}_{ (1/2){\bf{k+q/2}} }({\bf{k+q/2}})(t)
d_{ (1/2){\bf{k-q/2}} }({\bf{k-q/2}})(t)
\end{equation}
\begin{equation}
d_{ -(1/2){\bf{q}} }(-{\bf{q}})(t) 
 = A_{ {\bf{q}} }f_{ -{\bf{q}} }e^{-i\mbox{ }\omega_{ {\bf{q}} }t}
 - B_{ {\bf{q}} }f^{\dagger}_{ {\bf{q}} }e^{i\mbox{ }\omega_{ {\bf{q}} }t}
\end{equation}
\begin{equation}
d^{\dagger}_{ (1/2){\bf{q}} }({\bf{q}})(t)
 = A_{ {\bf{q}} }f^{\dagger}_{ {\bf{q}} }e^{i\mbox{ }\omega_{ {\bf{q}} }t}
 - B_{ {\bf{q}} }f_{ -{\bf{q}} }e^{-i\mbox{ }\omega_{ {\bf{q}} }t}
\end{equation}
\begin{equation}
d_{ (1/2){\bf{k-q/2}} }({\bf{k-q/2}})(t)
 = A_{ {\bf{k-q/2}} }f_{ {\bf{k-q/2}} }e^{-i\mbox{ }\omega_{ {\bf{k-q/2}} }t}
 - B_{ {\bf{k-q/2}} }f^{\dagger}_{ {\bf{-k+q/2}} }
e^{i\mbox{ }\omega_{ {\bf{k-q/2}} }t}
\end{equation}
\begin{equation}
d^{\dagger}_{ (1/2){\bf{k+q/2}} }({\bf{k+q/2}})(t)
 = A_{ {\bf{k+q/2}} }f^{\dagger}_{ {\bf{k+q/2}} }
e^{i\mbox{ }\omega_{ {\bf{k+q/2}} }t}
 - B_{ {\bf{k+q/2}} }f_{ {\bf{-k-q/2}} }e^{-i\mbox{ }\omega_{ {\bf{k+q/2}} }t}
\end{equation}
Now define,
\[
S^{>}({\bf{q}}t) = \langle \rho_{ {\bf{q}} }(t) \rho_{ -{\bf{q}} }(0)\rangle
 = N\langle [ d_{ -(1/2){\bf{q}} }(-{\bf{q}})(t)
 + d^{\dagger}_{ (1/2){\bf{q}} }({\bf{q}})(t) ]
[ d_{ (1/2){\bf{q}} }({\bf{q}})(0)
 + d^{\dagger}_{ -(1/2){\bf{q}} }(-{\bf{q}})(0) ] \rangle
\]
\[
\sum_{ {\bf{k}}, {\bf{k}}^{'} }
\langle
 d^{\dagger}_{ (1/2){\bf{k+q/2}} }({\bf{k+q/2}})(t)
 d_{ (1/2){\bf{k-q/2}} }({\bf{k-q/2}})(t)
 d^{\dagger}_{ (1/2){\bf{k^{'}-q/2}} }({\bf{k^{'}-q/2}})(0)
 d_{ (1/2){\bf{k^{'}+q/2}} }({\bf{k^{'}+q/2}})(0)
\rangle
\]
\[
= N
(\frac{ \epsilon_{ {\bf{q}} } }{\omega_{ {\bf{q}} } })
 \mbox{ }exp(-i \mbox{ }\omega_{ {\bf{q}} }t)
\]
\[
+ \sum_{ {\bf{k}}, {\bf{k}}^{'} }
\langle B_{ {\bf{k+q/2}} }
f_{ {\bf{-k-q/2}} }e^{-i\omega_{ {\bf{k+q/2}} }t}
A_{  {\bf{k-q/2}} }f_{ {\bf{k-q/2}} }e^{-i\omega_{ {\bf{k-q/2}} }t}
A_{  {\bf{k^{'}-q/2}} }f^{\dagger}_{ {\bf{k^{'}-q/2}} }
B_{ {\bf{k^{'}+q/2}} }f^{\dagger}_{ {\bf{-k^{'}-q/2}} } \rangle
\]
\[
+ \sum_{ {\bf{k}}, {\bf{k}}^{'} }
\langle B_{ {\bf{k+q/2}} }
f_{ {\bf{-k-q/2}} }e^{-i\omega_{ {\bf{k+q/2}} }t}
B_{  {\bf{k-q/2}} }f^{\dagger}_{ {\bf{-k+q/2}} }e^{i\omega_{ {\bf{k-q/2}} }t}
B_{  {\bf{k^{'}-q/2}} }f_{ {\bf{-k^{'}+q/2}} }
B_{ {\bf{k^{'}+q/2}} }f^{\dagger}_{ {\bf{-k^{'}-q/2}} } \rangle
\]
\[
S^{>}({\bf{q}},t) = N
(\frac{ \epsilon_{ {\bf{q}} } }{\omega_{ {\bf{q}} } })
 \mbox{ }exp(-i \mbox{ }\omega_{ {\bf{q}} }t)
\]
\[
+ \sum_{ {\bf{k}} }
 exp(-i \mbox{ }(\omega_{ {\bf{k+q/2}} }+\omega_{ {\bf{k-q/2}} })t)
[ B^{2}_{ {\bf{k+q/2}} }A^{2}_{ {\bf{k-q/2}} } + 
 B_{ {\bf{k+q/2}} }B_{ {\bf{-k+q/2}} }A_{ {\bf{k-q/2}} }
A_{ {\bf{k+q/2}} } ]
\]
\[
S^{<}({\bf{q}},t) =  N 
(\frac{ \epsilon_{ {\bf{q}} } }{\omega_{ {\bf{q}} } })
\mbox{ }exp(i \mbox{ }\omega_{ {\bf{q}} }t)
+ \sum_{ {\bf{k}} }
 exp(i \mbox{ }(\omega_{ {\bf{k+q/2}} }+\omega_{ {\bf{k-q/2}} })t)
[ B^{2}_{ {\bf{k-q/2}} }A^{2}_{ {\bf{k+q/2}} } +
 B_{ {\bf{k-q/2}} }B_{ {\bf{-k-q/2}} }A_{ {\bf{k+q/2}} }
A_{ {\bf{k-q/2}} } ]
\]
and, 
\[
 S^{>}({\bf{q}},t) = \langle \rho_{ {\bf{q}} }(t)  \rho_{ -{\bf{q}} }(0) \rangle
\]
\[
 S^{<}({\bf{q}},t) = \langle \rho_{ -{\bf{q}} }(0)  \rho_{ {\bf{q}} }(t) \rangle
\]
From the above equations, it is easy to see that there is a coherent
 part corresponding to the Bogoliubov spectrum and an incoherent part which is
 due to correlations and is responsible (hopefully) for the
 roton minimum.

\section{Acknowledgements}
 The author is deeply indebted to Prof. A. J. Leggett
 and Prof. A. H. Castro-Neto for very illuminating conversations and
 indeed implicitly hinting at the next step in the developments. 
 They should have been the coauthors but are not due to their own
 choosing.

\end{document}